\documentclass[usegraphicx,useAMS,usenatbib]{mn2e}

\usepackage{amssymb}
\usepackage{times}
\usepackage{aas_macros} 

\newcommand{\fridlund}{Bruntt, Fridlund \& Deleuil (2010, A\&A, submitted)}
\newcommand{\fridlundalt}{Bruntt, Fridlund \& Deleuil 2010, A\&A, submitted}

\newcommand{\dens}{$\rho$}
\newcommand{\rrad}{$R$}

\newcommand{\ione}{\,{\sc i}}
\newcommand{\itwo}{\,{\sc ii}}

\newcommand{\vald}{{VALD}}
\newcommand{\vwa}{{VWA}}

\newcommand{\str}{Str\"omgren}

\newcommand{\vmacrooo}{{$v_{\rm mac}$}}
\newcommand{\vmicrooo}{{$v_{\rm mic}$}}
\newcommand{\vmacroo}{{$v_{\rm macro}$}}
\newcommand{\vmicroo}{{$v_{\rm micro}$}}
\newcommand{\vmacro}{{macroturbulence}}
\newcommand{\vmicro}{{microturbulence}}
\newcommand{\largesep}{$\Delta \nu$}
\newcommand{\numax}{$\nu_{\rm max}$}

\newcommand{\muhz}{$\mu$Hz}

\newcommand{\teffsp}{$T_{\rm eff}^\star$}
\newcommand{\teff}{$T_{\rm eff}$}
\newcommand{\teffs}{$T_{\rm eff}$s}

\newcommand{\logg}{$\log g$}

\newcommand{\fehsp}{[Fe/H]$^\star$}
\newcommand{\feh}{[Fe/H]}

\newcommand{\vsini}{$v \sin i$}
\newcommand{\kms}{km\,s$^{-1}$}
\newcommand{\kmss}{[km/s]}

\newcommand{\msun}{$M_\odot$}
\newcommand{\fbol}{$f_{\rm bol}$}
\newcommand{\thetald}{$\theta_{\rm LD}$}

\newcommand{\angdim}{$\theta$}

\newcommand{\ie}{i.e.}
\newcommand{\cf}{cf.}
\newcommand{\eg}{e.g.}

\newcommand{\parallax}{$\pi_{\rm p}$}

\newcommand{\percent}{\,\%}

\newcommand{\chisq}{$\chi^2$}
\newcommand{\mgone}{Mg\,{\sc i}b}
\newcommand{\naone}{Na\,{\sc i}\,D}
\newcommand{\feonetwo}{Fe{\sc i}$+${\sc ii}}
\newcommand{\feone}{Fe{\sc i}}

\newcommand{\nione}{Ni{\sc i}}

\newcommand{\flux}{\,nW\,m$^{-2}$}

\newcommand{\wire}{{\em WIRE}}
\newcommand{\most}{{\em MOST}}
\newcommand{\corot}{{\em CoRoT}}
\newcommand{\cobe}{{\em COBE}}
\newcommand{\kepler}{{\em Kepler}}

\newcommand{\song}{{SONG}}
\newcommand{\ucles}{{UCLES}}
\newcommand{\harps}{{HARPS}}

\newcommand{\fies}{{FIES}}
\newcommand{\hipp}{{\em Hipparcos}} 
\newcommand{\chara}{{CHARA}}
\newcommand{\pavo}{{PAVO}}
\newcommand{\amber}{{AMBER}}
\newcommand{\vlti}{{VLTI}}
\newcommand{\susi}{{SUSI}}

\newcommand{\taupsa}{$\tau$~PsA}
\newcommand{\gammapav}{$\gamma$~Pav}
\newcommand{\etaser}{$\eta$~Ser}
\newcommand{\hdcarrier}{HR~5803} 
\newcommand{\pup}{171~Pup}
\newcommand{\ksihya}{$\xi$~Hya}
\newcommand{\betahyi}{$\beta$~Hyi}
\newcommand{\procyon}{Procyon\,A}
\newcommand{\acena}{$\alpha$~Cen\,A}
\newcommand{\acenb}{$\alpha$~Cen\,B}
\newcommand{\acenab}{$\alpha$~Cen\,A$+$B}
\newcommand{\muara}{$\mu$~Ara}
\newcommand{\nuind}{$\nu$~Ind}
\newcommand{\etaboo}{$\eta$~Boo}
\newcommand{\taucet}{$\tau$~Cet}
\newcommand{\tauceti}{$\tau$~Cet}
\newcommand{\deltaeri}{$\delta$~Eri}
\newcommand{\deltapav}{$\delta$~Pav}
\newcommand{\alphafor}{$\alpha$~For}

\newcommand{\betaaql}{$\beta$~Aql}
\newcommand{\gammaser}{$\gamma$~Ser}
\newcommand{\iotahor}{$\iota$~Hor}
\newcommand{\opha}{70~Oph\,A}
\newcommand{\betavir}{$\beta$~Vir}
\newcommand{\alfmen}{$\alpha$~Men}
\newcommand{\alphamen}{$\alpha$~Men}

\title[Accurate fundamental parameters for 23 bright solar-type stars]{Accurate fundamental parameters for 23 bright solar-type stars}

\author[H. Bruntt et al.]  
         {H.\ Bruntt$^{1,2}$\thanks{E-mail:bruntt@phys.au.dk},
          T.\ R.\ Bedding$^{2}$,
          P.-O.\ Quirion$^{3,4}$,
          G.\ Lo~Curto$^{5}$,
          F.\ Carrier$^{6}$,
          B.\ Smalley$^{7}$,\newauthor
          T.\ H. Dall$^{8}$,
          T.\ Arentoft$^{4}$,
          M.\ Bazot$^{9}$,
          R.\ P.\ Butler$^{10}$\\
$^{1}$LESIA, Observatoire de Paris-Meudon, 92195, France\\
$^{2}$Sydney Institute for Astronomy, School of Physics, The University of Sydney, 2006 NSW, Australia\\
$^{3}$Canadian Space Agency, 6767 Boulevard de l'A\'eroport, Saint-Hubert, Qu\'ebec J3Y 8Y9, Canada\\
$^{4}$Danish AsteroSeismology Centre (DASC), Department of Physics and Astronomy, University of Aarhus, DK-8000 Aarhus C, Denmark\\
$^{5}$European Southern Observatory, Alonso de Cordova 3107, Vitacura, Santiago, Chile\\
$^{6}$Instituut voor Sterrenkunde, Katholieke Universiteit Leuven, Belgium\\
$^{7}$Astrophysics Group, Keele University, Keele, Staffordshire ST5~5BG, UK\\
$^{8}$European Southern Observatory, Karl Schwarzschild Str.~2, 85748 Garching bei M\"unchen, Germany\\
$^{9}$Universidade do Porto, Centro de Astrof\'isica, Rua das Estrelas, PT 4150-762 Porto, Portugal\\
$^{10}$Carnegie Institution of Washington, Department of Terrestrial Magnetism, 5241 Broad Branch Road NW, Washington, DC 20015-1305, USA
}
\begin{document}
\date{Accepted XXX February 2010. Received ZZZZ December 2009}
\pagerange{\pageref{firstpage}--\pageref{lastpage}} \pubyear{2010}
\maketitle
\label{firstpage}
\begin{abstract}
We combine results from interferometry, asteroseismology
and spectroscopy to determine accurate fundamental parameters of 23 bright solar-type stars, from 
spectral type F5 to K2 and luminosity classes III to V.
For some stars we can use direct techniques to determine 
the mass, radius, luminosity and effective temperature,
and we compare with indirect methods that rely on photometric calibrations or spectroscopic analyses.
We use the asteroseismic information available in the literature 
to infer an indirect mass with an accuracy of 4--15\percent. 
From indirect methods we determine luminosity and radius to 3\percent.
We find evidence that the luminosity from the indirect method 
is slightly overestimated ($\approx 5$\percent) for the coolest stars,
indicating that their bolometric corrections are too negative.
For \teff\ we find a slight offset of $-40\pm20$~K between the spectroscopic method
and the direct method, meaning the spectroscopic temperatures are too high. 
From the spectroscopic analysis
we determine the detailed chemical composition for 13 elements, including Li, C and O.
The metallicity ranges from ${\rm [Fe/H]}=-1.7$ to $+0.4$, 
and there is clear evidence for $\alpha$-element enhancement in the metal-poor stars.
We find no significant offset between the spectroscopic surface gravity and 
the value from combining asteroseismology with radius estimates.
From the spectroscopy we also determine \vsini\ and 
we present a new calibration of macro- and microturbulence.
From the comparison between the results from the direct and spectroscopic methods we
claim that we can determine \teff, \logg, and \feh\ with absolute accuracies of 80~K, 0.08~dex, and 0.07~dex.
Photometric calibrations of \str\ indices provide accurate results for \teff\ and \feh\ but 
will be more uncertain for distant stars when interstellar reddening becomes important.
The indirect methods are important to obtain reliable estimates of the fundamental parameters
of relatively faint stars when interferometry cannot be used.
Our study is the first to compare direct and indirect methods for a large sample of stars,
and we conclude that indirect methods are valid, although slight corrections may be needed.
\end{abstract}





\begin{figure} 
\begin{center} 
\includegraphics[width=8.8cm]{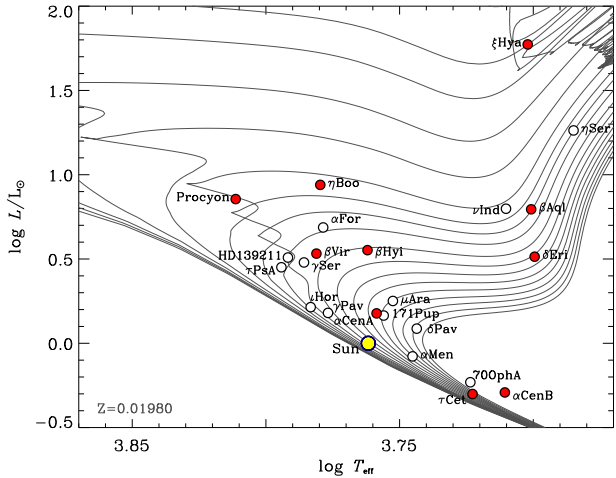} 
\caption{Hertzsprung-Russell diagram with the observed targets. 
Filled circles are used for stars with a measured angular diameter.
Evolution tracks from ASTEC are shown (metallicity $Z=0.0198$, mixing length $\alpha=1.8$) 
and the location of the Sun is shown for reference.
\label{fig:hr} }
\end{center} 
\end{figure} 

\section{Introduction}

Rapid progress in the fields of exoplanet science and asteroseismology has given
renewed importance to the task of determining fundamental stellar parameters.
The two main parameters that describe the evolution and structure
of a star are mass and chemical composition. 
Additional stellar parameters include the radius, luminosity, and rotation rate,
all of which change during the evolution of the star. 
The composition of a star will also change during its evolution due to
fusion in the core, mixing from rotation, overshooting and diffusion.
In later evolutionary stages, the effects of deep convection 
dredge-up and mass loss become important.
The fundamental parameters of stars can be determined using
various techniques like stellar interferometry 
(to obtain the radius and effective temperature)
and radial velocities and light curves of detached eclipsing binary stars 
(to get mass and radius).
These methods are important 
since they are nearly independent of assumptions about the stars themselves 
(the exception is limb darkening but the dependence is weak).
However, the methods mentioned here can only be applied to a few of the brightest and nearest stars.
For more distant stars we have to rely on calibration of photometric indices
or atmospheric models to infer the fundamental parameters from spectroscopic analyses. 
It is therefore of paramount importance to calibrate these indirect methods
using the model-independent methods for a significant sample 
of ``fundamental stars'' spanning a relatively wide range of spectral types.

Fundamental parameters are important in many areas of astrophysics and we
will briefly mention here the field of exoplanets and asteroseismology.
Characterization of the host stars of transiting exoplanets is essential since the
transit event only gives a measure of the relative radii of the planet and the star.
\cite{vanbelle09} used interferometry to measure the radii of 12 bright stars with known transiting planets. 
However, such methods are not available for the faint systems 
discovered by the \corot\ and \kepler\ missions.
In such cases, we have to rely on photometric calibrations
and spectroscopic analyses. To test different planet-formation scenarios and to understand
the diversity of planet systems, a full characterization is needed \citep{ammler09}.
The results of such investigations rest on the assumption that the spectroscopic methods 
can be applied. The true level of accuracy on the fundamental parameters must
therefore be systematically assessed for the entire range of spectral types of stars
hosting exoplanets.

Accurate fundamental parameters and associated realistic uncertainties 
are also essential input when studying the interior of stars 
through asteroseismic techniques. This is the only technique available
to probe the physical conditions inside the stars. 
Helioseismology has given us very detailed knowledge about the interior 
of the Sun (\eg, the depth of the outer convection zone, helium content, and interior rotation rate).
In the past 15 years, ground-based efforts using multi-site spectroscopy \citep{bedding08, aerts08}
and more recently photometric space missions like \wire, \most, \corot, and \kepler\
are allowing such studies of solar-like stars with a wide range of masses, 
chemical compositions and evolutionary stages. 
From the point of view of theoretical modelling it was shown by 
\cite{basu09} and \cite{stello09} that radii can be constrained 
to a few percent for the \kepler\ targets. 
This is an important result for the stars that host exoplanets in
order to get the absolute planetary radius \citep{jcd10}.
\cite{basu09} demonstrated that, in addition to asteroseismic measurements,
the effective temperature is an essential constraint. 
This can only be measured from well-calibrated photometry (which is affected by interstellar reddening) or spectroscopy.
For sub-giants and giant stars, \cite{basu09} pointed out that the metallicity is especially important.
Thus, the science output from \kepler\ relies on indirect methods.

In the current work we will make a homogeneous determination 
of the fundamental parameters of 23 solar-type stars with spectral types F5 to K2.
These stars are all bright and are good targets for ground-based asteroseismic campaigns.
Their locations in the Hertzsprung-Russell diagram are shown in Fig.~\ref{fig:hr}.
Also shown are evolution tracks from the ASTEC grid for metallicity $Z=0.0198$ \citep{jcd08}.
All except one star (\alfmen) have already been studied to varying extents
through asteroseismic techniques from the ground. 
The sample includes ten stars for which angular diameters have been measured
(marked by filled circles in Fig.~\ref{fig:hr}). 
The observed properties and derived stellar parameters are given in Tables~\ref{tab:fundall} and \ref{tab:fund}.

Ultimately, one would like to determine the fundamental 
parameters of stars using {\em direct methods} that are independent 
of model atmospheres or stellar evolution models. 
The mass, radius, luminosity and \teff\ can be determined 
in this way for some of the stars in our sample. 
For the remaining stars we need to rely on {\em indirect methods} 
that rely on model atmospheres (spectroscopic analysis)
or calibration of photometric indices. 

A model-independent mass can only be determined for stars 
in a binary system and this has been done in previous studies 
for four of the stars in our sample.
The mass can also be inferred indirectly by using asteroseismic measurements
and we will compare the two methods in Sect.~\ref{sec:mass}.

By combining measured parallaxes, angular diameters and bolometric fluxes 
(Sect.~\ref{sec:meas}) we can determine important 
global stellar parameters nearly independently of models: 
the radius, luminosity, and effective temperature (Sect.~\ref{sec:param}).
The determinations are slightly model-dependent, 
since the limb-darkening parameter depends on the adopted model atmosphere. 
The angular diameter has been measured for ten stars in the sample using interferometry.
We compare the results of these direct methods with \teff\ from
the calibration of \str\ indices and spectroscopic analysis. 
The comparison is used to quantify the applicability of the indirect methods,
which are the only options when analysing stars for which
parallaxes and angular diameters are not available. 
The composition of stars is an important ingredient for
the modelling of stars. In Sect.~\ref{sec:vwa} we make a detailed 
spectroscopic study to determine the effective temperature, 
surface gravity, chemical composition, and projected rotational velocity.

\begin{table*}
 \centering
  \caption{Observable properties of the target stars. $V$ is from SIMBAD and
parallaxes are all from from van Leeuwen (2007) except \acenab\ which is from S\"oderhjelm~(1999).
The \fbol\ is determined in Sect.~\ref{sec:fbol}.
The BC is from VandenBerg \& Clem (2003) and have an uncertainty of $0.03$~mag.
For four stars \largesep\ has not been measured and we give approximate values found using Eq.~\ref{eq:numax}
(indicated by the $\approx$ symbol).
References for the angular diameters and the asteroseismic data 
are given in the last two columns and explained in detail below the Table.
\label{tab:fundall}}
   \begin{tabular}{lrrrrrccrcc}
   \hline
       &   &     & $\pi_{\rm p}$ & \multicolumn{1}{c}{\thetald} &  \multicolumn{1}{c}{\fbol}          &    & \numax  &  \largesep    & Ref.     & Ref.        \\
Star   &HD & $V$ & [mas]         &   \multicolumn{1}{c}{[mas]}   &  \multicolumn{1}{c}{[nW\,m$^{-2}$]} & BC & [mHz]     & [$\mu$Hz]   & \thetald & Seis. \\  \hline


   \betahyi &   2151  &$ 2.80$&$134.07\pm 0.11$&$ 2.257\pm0.019$&$ 1.970\pm0.073$&$-0.07$&$ 1.00$ &$ 57.5 $ & 1    &i  \\ 
    \taucet &  10700  &$ 3.49$&$273.96\pm 0.17$&$ 2.022\pm0.011$&$ 1.140\pm0.038$&$-0.18$&$ 4.50$ &$169.0 $ & 2    &l  \\
   \iotahor &  17051  &$ 5.41$&$ 58.24\pm 0.22$&$              $&$              $&$-0.02$&$ 2.70$ &$120.0 $ &      &m  \\
  \alphafor &  20010  &$ 3.85$&$ 70.24\pm 0.45$&$              $&$              $&$-0.05$&$ 1.10$ &$ \approx 57$&  &i  \\
  \deltaeri &  23249  &$ 3.51$&$110.62\pm 0.22$&$ 2.394\pm0.029$&$ 1.180\pm0.046$&$-0.26$&$ 0.70$ &$ 43.8 $ & 3    &b  \\
  \alphamen &  43834  &$ 5.09$&$ 98.05\pm 0.14$&$              $&$              $&$-0.10$&$     $ &$      $ &      &   \\
   \procyon &  61421  &$ 0.34$&$284.52\pm 1.27$&$ 5.446\pm0.031$&$17.600\pm0.483$&$+0.00$&$ 1.00$ &$ 55.0 $ & 4    &j  \\
       \pup &  63077  &$ 5.37$&$ 65.75\pm 0.51$&$              $&$              $&$-0.12$&$ 2.05$ &$ 97.1 $ &      &d  \\
    \ksihya & 100407  &$ 3.55$&$ 25.14\pm 0.16$&$ 2.386\pm0.021$&$ 1.170\pm0.046$&$-0.24$&$ 0.09$ &$  6.8 $ & 3    &g  \\
   \betavir & 102870  &$ 3.63$&$ 91.50\pm 0.22$&$ 1.450\pm0.018$&$ 0.915\pm0.032$&$-0.02$&$ 1.40$ &$ 72.1 $ & 5    &e  \\
    \etaboo & 121370  &$ 2.70$&$ 87.77\pm 1.24$&$ 2.204\pm0.011$&$ 2.140\pm0.063$&$-0.01$&$ 0.75$ &$ 39.9 $ & 6    &h  \\
     \acena & 128620  &$ 0.00$&$747.10\pm 1.20$&$ 8.511\pm0.020$&$26.300\pm0.900$&$-0.06$&$ 2.40$ &$106.2 $ & 7    &i  \\
     \acenb & 128621  &$ 1.33$&$747.10\pm 1.20$&$ 6.001\pm0.034$&$ 8.370\pm0.349$&$-0.22$&$ 4.10$ &$161.4 $ & 7    &i  \\
 \hdcarrier & 139211  &$ 5.95$&$ 32.32\pm 0.50$&$              $&$              $&$-0.02$&$ 1.80$ &$ 85.0 $ &      &d  \\
  \gammaser & 142860  &$ 3.85$&$ 88.85\pm 0.18$&$              $&$              $&$-0.04$&$ 1.60$ &$ \approx 79$&  &i  \\
     \muara & 160691  &$ 5.15$&$ 64.48\pm 0.31$&$              $&$              $&$-0.08$&$ 2.00$ &$ 90.0 $ &      &c  \\
      \opha & 165341  &$ 4.03$&$196.66\pm 0.84$&$              $&$              $&$-0.17$&$ 4.50$ &$161.7 $ &      &f  \\
    \etaser & 168723  &$ 3.26$&$ 53.93\pm 0.18$&$              $&$              $&$-0.33$&$ 0.13$ &$  8.0 $ &      &a  \\
   \betaaql & 188512  &$ 3.70$&$ 73.00\pm 0.20$&$ 2.180\pm0.090$&$ 0.979\pm0.032$&$-0.25$&$ 0.41$ &$ \approx 26$ &8&i  \\
  \deltapav & 190248  &$ 3.56$&$163.71\pm 0.17$&$              $&$              $&$-0.10$&$ 2.30$ &$ \approx 106$& &i  \\
  \gammapav & 203608  &$ 4.22$&$107.98\pm 0.19$&$              $&$              $&$-0.09$&$ 2.60$ &$120.4 $ &      &k  \\
    \taupsa & 210302  &$ 4.95$&$ 54.70\pm 0.28$&$              $&$              $&$-0.01$&$ 1.95$ &$ 89.5 $ &      &d  \\
     \nuind & 211998  &$ 5.29$&$ 34.84\pm 0.27$&$              $&$              $&$-0.25$&$ 0.32$ &$ 25.1 $ &      &i  \\


\hline
\multicolumn{11}{l}{{\em References for} \thetald:}\\
\multicolumn{11}{l}{1=\cite{north07}; 2=\cite{teixeira09}; 3=\cite{thevenin05};4=\cite{kervella04}, \cite{mozur03},}\\
\multicolumn{11}{l}{and \cite{nordgren01}; 5=\cite{north09}; 6=\cite{thevenin05}, \cite{mozur03}, \cite{nordgren01}, }\\
\multicolumn{11}{l}{and \cite{vanbelle07}; 7=\cite{kervella03}; 8=\cite{nordgren99}.}\\

\\
\multicolumn{11}{l}{{\em References for} \numax\ {\em and} \largesep:}\\
\multicolumn{11}{l}{a=\cite{barban04}; b=\cite{bouchy03}; c=\cite{bouchy05}; d=F.~Carrier (private communication);}\\
\multicolumn{11}{l}{e=\cite{carrier05}, f=\cite{carrier06}; g=\cite{frandsen02}; h=\cite{carrier05};}\\
\multicolumn{11}{l}{i=\cite{kjeldsen08amp}; j=\cite{arentoft08}; k=\cite{mosser08}; l=\cite{teixeira09}; m=\cite{vauclair08}.}\\


\end{tabular}
\end{table*}
%

%
\begin{table*}
 \centering
  \caption{Fundamental parameters of the 23 targets as determined from 
the methods or parameters given in parenthesis: $\langle$\,$\rangle$,
\eg\ binary orbit, asteroseismic information ($\rho$), interferometry ($\theta_{\rm LD}$), 
parallax ($\pi_{\rm p}$), bolometric flux ($f_{\rm bol}$), and spectroscopic analysis with \vwa. 
Spectroscopic \teffs\ are offset by $-40$~K as discussed in the text.
The uncertainties on the spectroscopic and photometric \teffs\ are 80 and 90~K, respectively, 
and 0.07 and 0.10~dex for \feh.
Parameters from indirect methods are marked by the $\star$ symbol. 
\label{tab:fund}}
\setlength{\tabcolsep}{1.8pt} 
   \begin{tabular}{l|rr|rrr rr   cc  cc}
   \hline

         & 
\multicolumn{1}{c}{$M/{\rm M}_\odot$} & \multicolumn{1}{c}{$M^\star/{\rm M}_\odot$} & 
\multicolumn{1}{c}{$R/{\rm R}_\odot$} & \multicolumn{1}{c}{$R^\star/{\rm R}_\odot$} &
\multicolumn{1}{c}{$L/{\rm L}_\odot$} & \multicolumn{1}{c}{$L^\star/{\rm L}_\odot$} &
\multicolumn{1}{c}{\teff\ [K]} & \multicolumn{1}{c}{\teffsp\ [K]} & \multicolumn{1}{c}{\teffsp\ [K]} & 
\multicolumn{1}{c}{\fehsp} & \multicolumn{1}{c}{\fehsp} \\

Star & 
\multicolumn{1}{c}{$\langle$Binary orbit$\rangle$} & \multicolumn{1}{c}{$\langle$$\rho, R\rangle$} & 
\multicolumn{1}{c}{$\langle$$\theta_{\rm LD}, \pi_{\rm p}$$\rangle$} & \multicolumn{1}{c}{$\langle$$L, T_{\rm eff}$$\rangle$} &
\multicolumn{1}{c}{$\langle$$f_{\rm bol}, \pi_{\rm p}$$\rangle$} & \multicolumn{1}{c}{$\langle$ $V,{\rm BC},\pi_{\rm p}$  $\rangle$} &
\multicolumn{1}{c}{$\langle$$\theta_{\rm LD}, f_{\rm bol}$$\rangle$} &
\multicolumn{1}{c}{$\langle$Spec.$\rangle$} &
\multicolumn{1}{c}{$\langle$Phot.$\rangle$} &
\multicolumn{1}{c}{$\langle$Spec.$\rangle$} &
\multicolumn{1}{c}{$\langle$Phot.$\rangle$} \\


\hline
 \betahyi &$             $&$  ^a 1.08\pm0.05$&$ 1.810\pm0.015$&$ 1.89\pm0.06$&$ 3.41\pm0.13$&$ 3.57\pm0.11$&$5840\pm 59$&$5790$ &$5870$ & $-0.10$  & $-0.04$  \\
 \taucet  &$             $&$  ^a 0.79\pm0.03$&$ 0.794\pm0.005$&$ 0.85\pm0.03$&$ 0.47\pm0.02$&$ 0.50\pm0.02$&$5383\pm 47$&$5290$ &$5410$ & $-0.48$  & $-0.42$  \\
 \iotahor &$             $&$  ^c 1.23\pm0.12$&$              $&$ 1.16\pm0.04$&$            $&$ 1.64\pm0.05$&$          $&$6080$ &$6110$ & $+0.15$  & $-0.00$  \\
 \alphafor&$             $&$  ^d 1.53\pm0.18$&$              $&$ 2.04\pm0.06$&$            $&$ 4.87\pm0.16$&$          $&$6015$ &$6105$ & $-0.28$  & $-0.16$  \\
 \deltaeri&$             $&$  ^a 1.33\pm0.07$&$ 2.327\pm0.029$&$ 2.41\pm0.09$&$ 3.00\pm0.12$&$ 3.26\pm0.14$&$4986\pm 57$&$5015$ &       & $+0.15$  &          \\
 \alphamen&$             $&$                $&$              $&$ 0.99\pm0.03$&$            $&$ 0.83\pm0.03$&$          $&$5570$ &$5580$ & $+0.15$  & $+0.07$  \\
 \procyon &$1.461\pm0.025$&$  ^a 1.45\pm0.07$&$ 2.059\pm0.015$&$ 2.13\pm0.06$&$ 6.77\pm0.20$&$ 7.17\pm0.22$&$6494\pm 48$&$6485$ &$6595$ & $+0.01$  & $+0.02$  \\
 \pup     &$             $&$  ^c 0.99\pm0.11$&$              $&$ 1.24\pm0.04$&$            $&$ 1.46\pm0.05$&$          $&$5710$ &$5790$ & $-0.86$  & $-0.75$  \\
 \ksihya  &$             $&$  ^b 2.89\pm0.23$&$10.207\pm0.111$&$10.14\pm0.39$&$57.65\pm2.39$&$59.38\pm2.58$&$4984\pm 54$&$5045$ &       & $+0.21$  &          \\ 
 \betavir &$             $&$  ^a 1.42\pm0.08$&$ 1.704\pm0.022$&$ 1.69\pm0.05$&$ 3.40\pm0.12$&$ 3.40\pm0.10$&$6012\pm 64$&$6050$ &$6150$ & $+0.12$  & $+0.10$  \\
 \etaboo  &$             $&$  ^a 1.77\pm0.11$&$ 2.701\pm0.040$&$ 2.66\pm0.09$&$ 8.65\pm0.35$&$ 8.66\pm0.36$&$6028\pm 47$&$6030$ &$6025$ & $+0.24$  & $+0.27$  \\
 \acena   &$1.105\pm0.007$&$  ^a 1.14\pm0.05$&$ 1.225\pm0.004$&$ 1.24\pm0.04$&$ 1.47\pm0.05$&$ 1.51\pm0.05$&$5746\pm 50$&$5745$ &$5635$ & $+0.22$  & $+0.12$  \\
 \acenb   &$0.934\pm0.006$&$  ^a 0.92\pm0.04$&$ 0.864\pm0.005$&$ 0.91\pm0.03$&$ 0.47\pm0.02$&$ 0.51\pm0.02$&$5140\pm 56$&$5145$ &$    $ & $+0.30$  &          \\
\hdcarrier&$             $&$  ^c 1.52\pm0.16$&$              $&$ 1.56\pm0.05$&$            $&$ 3.23\pm0.14$&$          $&$6200$ &$6280$ & $-0.04$  & $-0.11$  \\
\gammaser &$             $&$  ^d 1.30\pm0.15$&$              $&$ 1.55\pm0.05$&$            $&$ 3.02\pm0.09$&$          $&$6115$ &$6245$ & $-0.26$  & $-0.19$  \\
 \muara   &$             $&$  ^c 1.21\pm0.13$&$              $&$ 1.39\pm0.05$&$            $&$ 1.78\pm0.06$&$          $&$5665$ &$5690$ & $+0.32$  & $+0.19$  \\
 \opha    &$0.890\pm0.020$&$  ^c 1.10\pm0.13$&$              $&$ 0.91\pm0.03$&$            $&$ 0.59\pm0.02$&$          $&$5300$ &$    $ & $+0.12$  &          \\
 \etaser  &$             $&$  ^d 1.45\pm0.21$&$              $&$ 6.10\pm0.25$&$            $&$18.32\pm0.87$&$          $&$4850$ &$    $ & $-0.11$  &          \\
 \betaaql &$             $&$  ^b 1.26\pm0.18$&$ 3.211\pm0.133$&$ 3.30\pm0.12$&$ 5.73\pm0.19$&$ 6.23\pm0.25$&$4986\pm111$&$5030$ &       & $-0.21$  &          \\
 \deltapav&$             $&$  ^d 1.07\pm0.13$&$              $&$ 1.20\pm0.04$&$            $&$ 1.22\pm0.04$&$          $&$5550$ &$5540$ & $+0.38$  & $+0.33$  \\
 \gammapav&$             $&$  ^c 1.21\pm0.12$&$              $&$ 1.15\pm0.04$&$            $&$ 1.52\pm0.05$&$          $&$5990$ &$6135$ & $-0.74$  & $-0.57$  \\
 \taupsa  &$             $&$  ^c 1.34\pm0.13$&$              $&$ 1.45\pm0.04$&$            $&$ 2.82\pm0.09$&$          $&$6235$ &$6385$ & $+0.01$  & $+0.05$  \\
 \nuind   &$             $&$  ^d 1.00\pm0.13$&$              $&$ 3.18\pm0.12$&$            $&$ 6.28\pm0.23$&$          $&$5140$ &$    $ & $-1.63$  &          \\


\hline
\multicolumn{12}{l}{Notes on column 3:
                   (a): Mean density, \dens, is scaled from \largesep; radius, \rrad, is from interferometry. 
                   (b): \dens\ is scaled from \numax; \rrad\ is from interferometry.}\\
\multicolumn{12}{l}{(c): \dens\ is scaled from \largesep; \rrad\ is from $R \propto L^{1/2} \, T_{\rm eff}^{-2}$.
                    (d): \dens\ is scaled from \numax; \rrad\ is from $R \propto L^{1/2} \, T_{\rm eff}^{-2}$.}

\end{tabular}
\end{table*}
%

\section{Stellar mass}
\label{sec:mass}

The mass of a single star is particularly difficult to determine.
We adopt an indirect method based on asteroseismic information 
and compare this with the fundamental method available for binary stars.
We will then compare our method with direct modelling approaches
found in the literature and discuss the accuracy that 
can be obtained on the masses of single stars.

\subsection{Mass from a binary orbit 
\label{sec:massbin}}

The absolute masses of stars in binary systems 
are determined from astrometric or radial velocity measurements by applying Kepler's laws.
This is possible for four of the stars in our sample: 
\procyon\ \citep{girard00, gatewood06},
\acena\ and~B \citep{pourbaix02}, and \opha\ \citep{eggenberger08}.
In Table~\ref{tab:fund} we list the measured masses (column~2). 
In the case of \procyon\ we give the weighted mean of the two studies
mentioned above. 
These four stars provide an important test for the indirect 
method for mass determination that we propose in the following.

\subsection{Mass from asteroseismic data
\label{sec:massast}}

\begin{figure} 
\begin{center} 
\includegraphics[width=9cm]{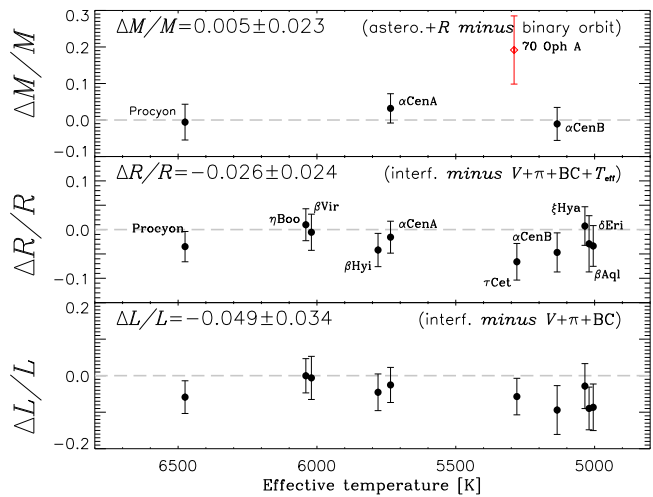} 
\caption{Relative differences between fundamental parameters as 
determined by direct and indirect methods (\cf\ Table~\ref{tab:fund}). 
In the top panel \opha\ is marked by an open symbol since 
its radius is determined from an indirect method and is therefore more uncertain.
The effective temperature on the abscissa is the spectroscopic value.
\label{fig:fund} }
\end{center} 
\end{figure}

The mass of a solar-type star can be inferred using 
asteroseismology (\eg\ \citealt{north07}).
The power spectrum of the observed time-series 
of radial velocities or a photometric light curve
is characterized by an envelope of excess power
consisting of a series of regular peaks (\eg\ Fig.~1 in \citealt{kjeldsen09}).
The oscillations are characterized by
the frequency of maximum power (\numax) and the large frequency separation ($\Delta \nu$).
Here, \largesep\ is the separation between peaks with the same spherical degree ($l$) and consecutive
radial order ($n$). In general, the large separation will vary slightly with frequency 
but its mean value scales with the square root of mean density of the star \citep{ulrich86}.
Measured relative to the solar value the expression is \citep{kjeldsen95}:
\begin{equation}
{\Delta \nu \over \Delta \nu_\odot} = 
\left({\rho\over \rho_{\odot}}\right)^{1/2} = \left({M\over {\rm M}_\odot}\right)^{1/2} \, \left({R\over {\rm R}_\odot}\right)^{-3/2}.
\label{eq:large}
\end{equation}
When the radius can be determined from interferometric measurements 
(combining angular diameter and parallax; see Sect.~\ref{sec:radius}) 
we can determine the mass from this expression. 
We have compared the mass for the binary stars and from Eq.~\ref{eq:large} 
for three stars (\procyon\ and \acenab) and the results agree very well,
as is shown in the upper panel in Fig.~\ref{fig:fund}. 
The mean relative difference is $\Delta M/M = +0.005 \pm 0.023$ (rms error).
We also plot the comparison for \opha\ (also in a binary system) but in 
this case there is no published angular diameter and so
the radius was derived from $R \propto L^{1/2} T_{\rm eff}^{-2}$ (see Sect.~\ref{sec:radius}).
As a consequence, the mass of \opha\ has a much larger uncertainty ($12$\percent) 
than the three other binary stars ($4.3$--$4.8$\percent).

For 18 stars in our sample, \largesep\ has been measured.
For four others, \alphafor, \gammaser, \betaaql\ and \deltapav, only \numax\ has been measured. 
However, \cite{kjeldsen95} determined a scaling between \largesep\ and \numax\ as
\begin{equation}
{\Delta \nu \over \Delta \nu_\odot} = 
\left({\nu_{\rm max} \over \nu_{{\rm max}; \odot}}\right)^{1/2}\, \left({R\over {\rm R}_\odot}\right)^{1/2}\, \left({T_{\rm eff}\over {\rm T}_{{\rm eff}; \odot}}\right)^{1/4}.
\label{eq:numax}
\end{equation}
We used this expression using the radius form Sect.~\ref{sec:radius}
and \teff\ from the spectroscopic analysis (Sect.~\ref{sec:vwa}).
A comparison with the more recent \largesep\,--\,\numax\ calibration of \cite{stello09numax} 
shows agreement within 1~$\sigma$ for all four stars.

In Table~\ref{tab:fundall} we give a list of the values of \largesep\ and \numax\ from the literature.
The quality and quantity of the data sets used for the determination of these asteroseismic values 
span a wide range. Some data sets are single-site observations lasting one night while 
in one case (\procyon), a multi-site campaign was carried out using 11 telescopes
lasting three weeks. In addition, some stars have been observed more than once.
It would be difficult to give a homogeneous set of uncertainties, 
so instead we adopted uncertainties on \largesep\ and \numax\ of 2 and 5\percent, respectively.
To determine the mass as described above, we used \largesep\ if it is available, otherwise we used \numax.
In the case of the two most evolved stars (\ksihya\ and \etaser) the large separation is 7--8\,\muhz.
It is well-known that it requires long time-series to be able
to resolve the individual peak for evolved stars. 
For this reason we adopted the more robust value of \numax\ for 
these two stars (see also \citealt{stello09}).

The determined masses of the 22 stars with asteroseismic information available 
are given in column~3 in Table~\ref{tab:fund}.
Four different combinations were used to determine the mean density (either Eq.~\ref{eq:large} or \ref{eq:numax})
and the radius (from interferometry or the relation $R\propto L^{-1/2}T_{\rm eff}^{-2}$; see Sect.~\ref{sec:radius}).
The combination used for each star is explained below Table~\ref{tab:fund}.
The propagation of errors yield uncertainties in the range 4--15\percent. 

We emphasize that our mass estimates rest on the assumption of the absolute 
validity of the scaling relations of Eq.~\ref{eq:large} and \ref{eq:numax}. 
The uncertainty estimate we give here comes only from the error propagation of 
$\Delta \nu$, $\nu_{\rm max}$ and $R$ (and also \teff\ for  Eq.~\ref{eq:numax}). 
The absolute values of the masses but also the uncertainties will require a more 
extensive study and the results given here should therefore be used with caution. 



\subsection{Accuracy of single-star masses}
\label{sec:massacc}

Using the same approach that we adopted in Sect.~\ref{sec:massast},
\cite{north07} analysed \betahyi\ and \cite{teixeira09} analysed \tauceti.
They obtained masses of $1.07\pm0.03$\,\msun\ and $0.783\pm0.012$\,\msun, respectively.
For \betahyi, \cite{kjeldsen08surf} corrected the asteroseismic data for surface effects
and used the updated \hipp\ parallax to obtain $1.085\pm0.028$\,\msun.
We used nearly the same data so it is not surprising that our values are almost identical:
$1.08\pm0.05$ and $0.79\pm0.03$\,\msun. The reason for the higher uncertainties
is that we assumed a common (larger) uncertainty on the large separation. 


We will now compare
our estimates with three studies that combine asteroseismic information 
(large separation or individual oscillation frequencies)
and classical constraints. In these studies, the classical constraints include a spectroscopic
\teff\ and metallicity and 
the luminosity is determined from the $V$ magnitude and parallax while
adopting a bolometric correction to obtain the bolometric flux 
(as described in Sect.~\ref{sec:lumn}).

The sub-giant \etaboo\ was analysed by \cite{guenther05} based on direct modelling
of the individual oscillation frequencies. They determined a mass of $1.71\pm0.05$\,\msun,
in good agreement with our value of $1.77\pm0.11$\,\msun.
The main sequence star \iotahor\ was modelled by \cite{vauclair08}.
They considered different values for the helium content and metallicity 
and determined a very precise mass $1.25\pm0.01$\,\msun, which agrees with our less precise
estimate of $1.15\pm0.12$\,\msun.
\cite{bedding06} determined a mass $0.85\pm0.04$\,\msun\ for the population-II sub-giant \nuind.
They constrained the location in the H-R diagram using
both classical constraints ($L/L = 6.21\pm0.23$ and $T_{\rm eff}=5300\pm100$\,K)
and asteroseismic information (the large separation). 
They used model grids for different overshoot parameters but with a fixed metallicity at $Z=0.001$. 
Our mass determined from the simple scaling is $1.05\pm0.12$\,\msun\ and 
is in agreement with \cite{bedding06}. 

The excellent agreement with the 4 stars with direct binary masses (top panel in Fig.~\ref{fig:fund}) 
and the good agreement for 3 detailed model-comparisons discussed 
above give us confidence in the method we adopted.
Although the uncertainties are relatively large, 
they offer a starting point for more detailed modelling, which can potentially lead to more
accurate estimates of the mass and age.  

\section{Direct measurements: \mbox{\boldmath \angdim}, \mbox{\boldmath \parallax} and \mbox{\boldmath \fbol}}
\label{sec:meas}

\subsection{Angular diameters
\label{sec:angular}}

The angular diameters of stars can be measured using interferometry and
can be used to infer nearly model-independent linear radii and effective
temperatures (see Sect.~\ref{sec:radius} and \ref{sec:teff}). 
The observations consist of measuring fringe visibility as a function of baseline.
The usual approach is to fit the observed visibilities with a uniformly 
illuminated disk and to convert the diameter to a limb-darkened (LD) disk using correction 
factors from atmospheric models. This is the approach used for the LD diameters 
taken from the literature and given in Table~\ref{tab:fund} for ten stars. 
However, we note that for more extended baselines with visibilities 
measured beyond the first null, a direct fit including
the limb-darkening is needed (e.g.\ as done for \procyon\ by \citealt{aufdenberg05}).


\cite{allende02} discussed the case of \procyon\ and in particular
the difference between plane-parallel 1D and 3D hydrodynamical models. 
They found that their 3D model results in a slightly smaller LD correction factor, resulting 
in smaller linear radii (1.6\percent) and thus a higher effective temperature (50~K).
\cite{bigot06} used a 3D model to infer the limb darkening parameter of the much
cooler star \acenb.
They also found a slightly smaller linear radius (0.3\percent) compared to their 1D model.
When comparing these results for 1D and 3D models it is 
important to note that \cite{allende02} used the angular diameter measured
in the visible \citep{mozur91} and \cite{bigot06} 
used infrared interferometric data. 
This may partly explain the difference between the studies (see also \citealt{aufdenberg05}), 
and future studies of other solar-type stars will shed more light on this issue.

For simplicity and uniformity we shall use the limb-darkened diameters (\thetald) from the literature
based on 1D~LTE models as originally published (for further discussion see Sect.~\ref{sec:teff})
and listed in Table~\ref{tab:fundall}.
In some cases more than one angular diameter exists and we have computed
the weighted mean value, as done by \cite{teixeira09} for \taucet.

\begin{figure} 
\begin{center} 
\includegraphics[width=8.8cm]{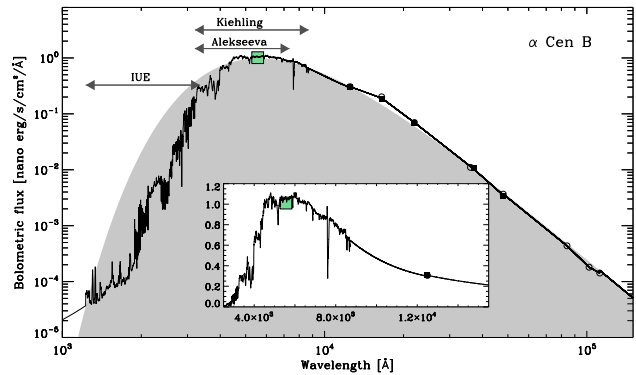} 
\caption{The bolometric flux of \acenb\ is determined 
by integrating the black curve shown here.
The ranges for the satellite IUE data and 
the spectrophotometric data are indicated by arrows and the name of the first author.
Box symbols and circles are broad-band fluxes.
The inset shows the details in the visual range on a linear scale.
\label{fig:fbol}}
\end{center} 
\end{figure} 

\subsection{Bolometric fluxes
\label{sec:fbol}}

The bolometric flux ($f_{\rm bol}$) can be estimated using UV measurements from
space, spectrophotometry in the visible region, 
and broadband magnitudes in the near infrared. 
Since the target stars are nearby we neglect interstellar 
absorption\footnote{All 23 stars are closer than 40~pc, and 
20 stars are closer than 20~pc.}.


We used UV spectrophotometry (1150--3350\,\AA) from 
the International Ultraviolet Explorer (IUE\footnote{The IUE spectra are available 
from http://sdc.laeff.inta.es/ines/.}).
Several dozen spectra are available for most stars except
\alfmen, \betavir, \etaboo, and \betaaql, where fewer than ten were available.
For \ksihya\ no IUE data are available and we used the broadband UV data from the 
Ultraviolet Sky Survey Telescope on the TD-1 satellite \citep{thompson78}.  
In the visual regime we used spectrophotometry
from \cite{breger76}, \cite{kiehling87} and \cite{aleks96}. 
In the near-IR and IR we used broadband photometry from \cite{morel78}, 
\cite{thomas73}, and \cite{engels81}. 
For the broadband data we calculated absolute fluxes 
using the values for Vega listed in \cite{cox00}. 
For a few targets we used additional flux values 
from the \cobe\ DIRBE point source catalog \citep{smith04}.

An example of the available data for \acenb\ is shown in Fig.~\ref{fig:fbol}. 
It comprises UV data from the IUE satellite, 
spectrophotometric data from \cite{kiehling87}
and broadband photometry from \cite{engels81} (squares)
and \cite{thomas73} (circles).
The grey shaded area is the Planck black body curve for the 
determined \teff\ after scaling by using the angular diameter.

The measured fluxes are listed in Table~\ref{tab:fundall} (column~6).
We find good agreement with previous determinations of \fbol\ in the literature.
For \betahyi, \cite{north07} found $2.019\pm0.050$\flux, compared to our value of $1.970\pm0.073$\flux.
For \betavir, \cite{north09} found $0.944\pm0.020$\flux\ and we have $0.915\pm0.032$\flux.
\cite{fuhrmann97} combined several estimates in the literature for \procyon\ to get a 
mean value of $18.20\pm0.43$\flux\ where we get $17.60\pm0.48$. 
For these three stars our \fbol\ values are all lower but agree 
with the literature values within $1\,\sigma$.


\subsection{\hipp\ parallaxes\label{sec:par}}

We used the updated \hipp\ parallaxes ($\pi_{\rm p}$) from \cite{leeuwen07}
and listed in Table~\ref{tab:fundall}.
The values agree well with the originally published results 
\cite{eso97} for all stars, but in one case the uncertainty is lower by a factor 4. 
This is for the giant star \ksihya\ where the original value is 
$\pi_{\rm p} = 72.95\pm0.83 $\,mas and the new value is 
$\pi_{\rm p} = 73.00\pm0.20$\,mas. 
This means that the uncertainty on the linear radius decreases 
significantly from $3.4$ to $1.0$\percent.
For the binary pair \acenab\ we used the parallax
from \cite{soderhjelm99} since this analysis takes into account the binary orbit.


\section{Indirect estimates of \mbox{\boldmath $R$}, \mbox{\boldmath $L$} and \mbox{\boldmath \teff}} 
\label{sec:param}

\subsection{Radius\label{sec:radius}}

The linear radius is determined from the limb darkened angular diameter 
and the parallax through the relation
\begin{equation}
R/{\rm R}_\odot = 9.30\times10^{-3}\, \theta_{\rm LD} / \pi_{\rm p},
\label{eq:radius}
\end{equation}
where we use values of \thetald\ and \parallax\ in milliarcseconds (mas) from Table~\ref{tab:fundall}.
The resulting linear radii are given in column~4 in Table~\ref{tab:fund}. 
The uncertainties on $R$ range from 0.2\percent\ (\acena) to 4\percent\ (\betaaql).

For 13 stars without measured angular diameters we estimated the radius from
the relation $R/R_\odot = (L/L_\odot)^{1/2} \, (T/T_{{\rm eff;}\,\odot})^{-2}$.
We have used indirect estimates for $L$ and \teff: 
$L$ is found using the $V$ magnitude, bolometric correction and parallax 
(Sect.~\ref{sec:lumn}) and \teff\ is from the spectroscopic analysis (Sect.~\ref{sec:teff}).
We obtain radii with a {\em precision} of about 2--4\percent\ from this method. 

The direct and indirect estimates of the radius are compared 
in the middle panel in Fig.~\ref{fig:fund}. There is very good 
agreement for all ten stars, with the largest deviation being $7\pm4$\percent\ for \taucet.
The mean relative difference of $\Delta R / R = -0.022 \pm 0.024$ (rms error) 
indicates that the claimed precision is realistic. 
Since we have a significant number of stars we propose that
the indirect method is {\em accurate} with a $1$-$\sigma$ uncertainty of $\approx 3$\percent.




\subsection{Luminosity\label{sec:lumn}}

The luminosity can be determined directly from the bolometric flux ($f_{\rm bol}$),
and the parallax ($\pi_{\rm p}$) through
\begin{equation}
L/{\rm L}_\odot = 3.12\times10^4\, f_{\rm bol}\, / \, \pi_{\rm p}^2,
\label{eq:lumn}
\end{equation}
where we insert the values of $f_{\rm bol}$ and $ \pi_{\rm p}$ from
Table~\ref{tab:fundall} in units of nW and mas, respectively. 
The resulting luminosities are given in column~6 in Table~\ref{tab:fund}.

In Sect.~\ref{sec:fbol} we described how we determined \fbol\ for ten stars.
The parallax is available from \cite{leeuwen07} for all targets.
We have compared the resulting $L$ with an indirect method that relies
on the apparent $V$ magnitude, bolometric correction 
(see Sect.~\ref{sec:bc}) and the parallax.
The bolometric magnitude is defined as $M_{\rm bol} = V + {\rm BC}_V + 5 \log \pi_{\rm p} + 5$.
We assumed that the interstellar reddening is zero since all targets are 
relatively close. 
We obtained $V$ from SIMBAD and adopted the uncertainty $\sigma(V) = 0.015$. 
We converted $M_{\rm bol}$ to $L/L_\odot$ using $M_{{\rm bol;}\odot} = 4.75\pm0.01$ (IAU recommendation 1999).

In the bottom panel in Fig.~\ref{fig:fund} the values are compared 
to the luminosity determined for ten stars combining the angular diameter with the parallax (Eq.~\ref{eq:lumn}).
The mean relative difference is $\Delta L/L = -0.048 \pm 0.033$ (rms error),
indicating there is little or no systematic error. 
The largest deviations are $9\pm6$\percent\ for the three cool 
stars \acenb, \betaaql\ and \deltaeri.
In fact, all five stars with \teff\ below 5300~K appear 
to have overestimated luminosities from the indirect method,
indicating that their BCs are too negative 
(Fig.~\ref{fig:bc} shows that BCs change rapidly for cool temperatures).
We note that adding $0.03$~mag to all BCs
would make the indirect luminosities and radii agree with 
the direct measures to within 1-$\sigma$ for all ten stars.
The mean relative differences would then improve to
$\Delta L/L = -0.020 \pm 0.034$ and $\Delta R / R = -0.010 \pm 0.027$.

From the comparison between the direct and indirect methods we conclude 
that the luminosity can be determined to $\approx 3$\percent\ from $V+\pi_{\rm p}+{\rm BC}$. 
The values are listed for all 23 stars in column~7 in Table~\ref{tab:fund}.
We caution that we find evidence for a slight overestimation of $L$ 
from the indirect method for cool stars.


\subsection{Bolometric correction
\label{sec:bc}}

\begin{figure} 
\begin{center} 
\includegraphics[width=8.8cm]{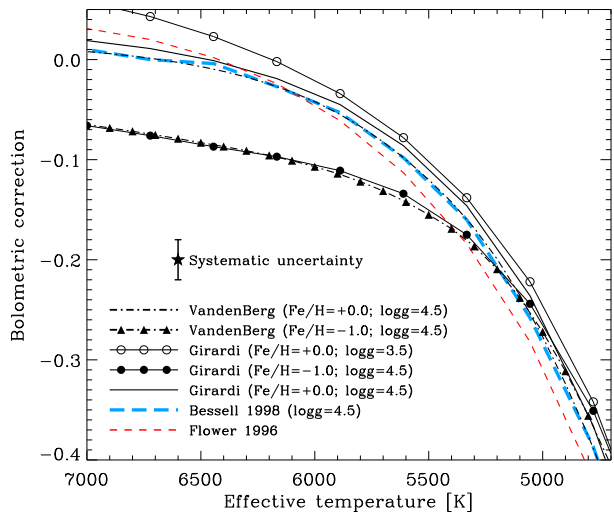} 
\caption{Bolometric corrections from four different sources. 
We have adopted the BCs from VandenBerg \& Clem (2003) 
that take into account the changes with metallicity and \logg.
\label{fig:bc} }
\end{center} 
\end{figure}

We have investigated the differences between some commonly used 
tabulations of bolometric corrections (BC), \ie\ from 
\cite{flower96}, \cite{bessell98}, \cite{girardi02} and \cite{clem03}.
For Girardi et al.\ we used the updated 
results from the dustyAGB07 database\footnote{{\em http://stev.oapd.inaf.it/dustyAGB07/}.}.
The BC values in these works are all based on atmospheric models and have
slightly different dependencies, as shown in Fig.~\ref{fig:bc}.
The BCs from \cite{flower96} only depend on \teff, 
while \cite{bessell98} values include a slight dependency with \logg.
Both \cite{clem03} and \cite{girardi02} include the changes 
in BC with \logg\ and \feh. This has an important impact on the BC for 
the metal-poor stars in our sample (\taucet, \pup, \gammapav\ and \nuind). 
In Fig.~\ref{fig:bc} we show the difference between 
${\rm [Fe/H]} = 0.0$ and $-1.0$ for \cite{girardi02} and \cite{clem03}.
For these two studies, the mean difference of the BCs 
for the 23 stars is negligible, $\Delta {\rm BC} = +0.007\pm0.019$ (rms error).

Inspecting the four tabulations of BC for ${\rm [Fe/H]}=0.0$ and $\log g=4.5$ in Fig.~\ref{fig:bc},
we find the maximum difference is about 0.04 mag in the range 5000--6500\,K.
Based on this we assign a 1-$\sigma$ systematic uncertainty of 0.02 mag to the BC.
We note that the BC tabulations we have compared depend on their adopted metallicity
scale relative to the Sun. However, recent studies of the metallicity 
of the Sun indicate a lower value than previously thought
(mostly due to revision of the light elements C and O; \citealt{caffau08}). 
This could mean that the opacities used for the calculation of the BCs include some bias,
especially above 5500~K, where Fig.~\ref{fig:bc} shows that metallicity plays an important role.

In the following we have adopted the tabulations of \cite{clem03} and 
the BC values are interpolated in their grids. We included the uncertainties 
on \teff\ (80~K), \logg\ (0.08~dex) and \feh\ (0.07~dex) to determine the BC 
for each star with an uncertainty of 0.03~mag.
The values for the atmospheric parameters (\teff, \logg\ and \feh) are taken from 
the spectroscopic analysis in Sect.~\ref{sec:vwa}.
The BC values are given in Table~\ref{tab:fundall} (column~7).

\begin{figure} 
\begin{center} 
\includegraphics[width=8.9cm]{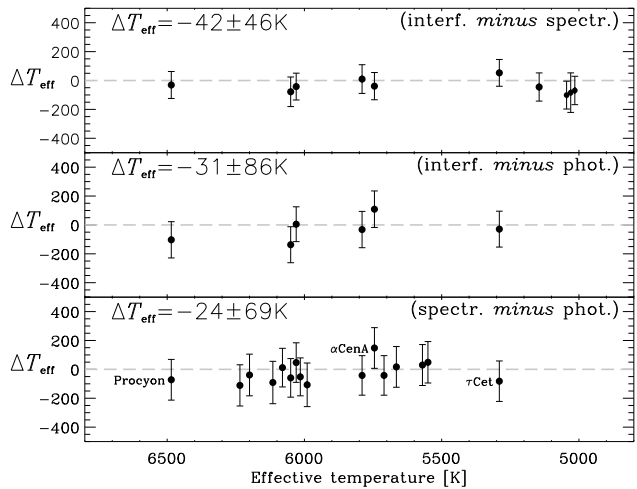}
\caption{Differences between \teff\
determined by direct and indirect methods (\cf\ Table~\ref{tab:fund}):
angular diameter and bolometric flux (''interf.''),
spectroscopic determination (``spectr.''),
and \str\ photometric indices (``phot.'').
The effective temperature on the abscissa is the spectroscopic value.
\label{fig:teff} }
\end{center} 
\end{figure} 




\begin{figure*} 
\begin{center} 
\includegraphics[angle=90,width=18.2cm]{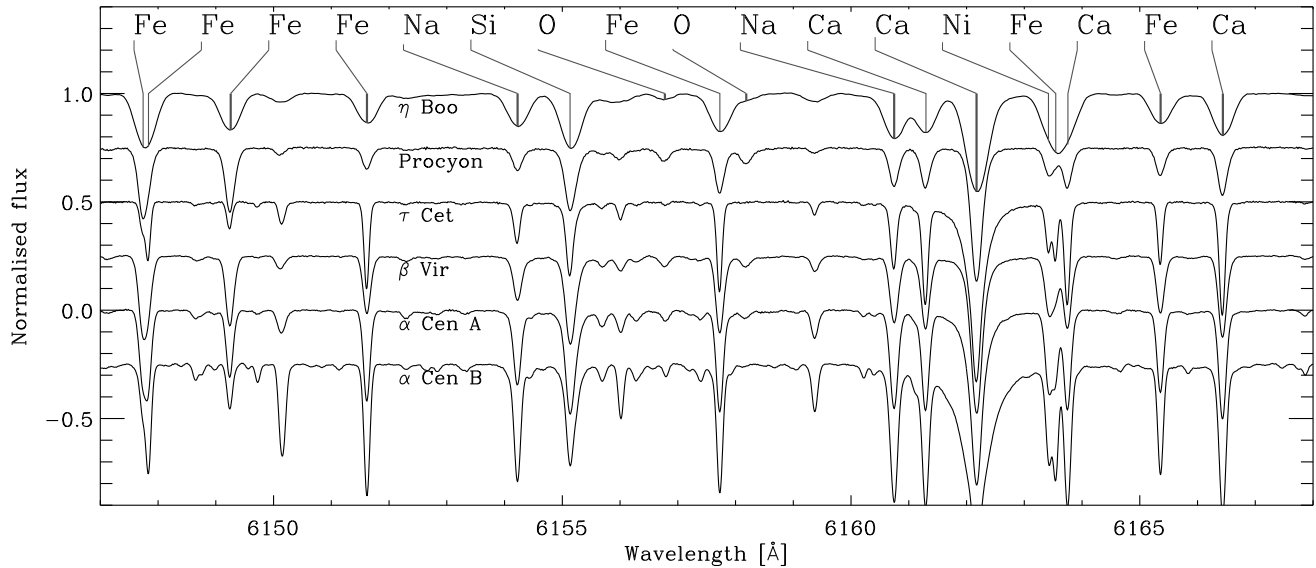}
\end{center} 
\caption{A small section of the observed spectra of six stars. 
These data represent 1\percent\ of the entire spectral ranged 
used to determine \teff, \logg, \vsini, and the chemical composition.
The wide Ca line at $\lambda6162$\,\AA\ is used to constrain the \logg\ value.
\label{fig:spec}}
\end{figure*} 


\subsection{Effective temperatures\label{sec:teff}}

The effective temperature can be directly determined 
from the limb-darkened angular diameter and the bolometric flux from the definition
\begin{equation}
\sigma \, T_{\rm eff}^4 = 4 \, f_{\rm bol} \, / \, \theta_{\rm LD}^2,
\label{eq:teffdef}
\end{equation}
where $\sigma$ is the Stefan-Boltzmann constant. 
Eq.~\ref{eq:teffdef} can also be expressed as 
\begin{equation}
T_{\rm eff} = 7402 \times f_{\rm bol}^{0.25}  / \theta_{\rm LD}^{0.5},
\label{eq:teff}
\end{equation}
where the unit of \fbol\ is nW and \thetald\ is in mas as given in Table~\ref{tab:fundall}.
In Table~\ref{tab:fund} we list the \teff\ determined for the ten stars
with measured angular diameters (column 8).

We used two indirect methods to determine \teff. 
One is based on detailed spectroscopic analysis of Fe\ione\ and \itwo\ lines 
using high-quality spectra and is described in Sect.~\ref{sec:vwa}.
The second method is based on \str\ photometric indices
with the recent calibration by \cite{holmberg07}. 
This calibration is not valid for the most evolved stars in the sample.
We used indices from the homogeneously calibrated catalogue of 
E.~H.~Olsen\footnote{The complete ``EHO'' catalogue of more than 30\,000 stars 
was obtained by private communication.} \citep{olsen94}.
We determined \teff\ from the \cite{holmberg07} calibration for 16 stars.
The spectroscopic and photometric \teffs\ are given in columns 9 and 10 in Table~\ref{tab:fund}.

The three methods are compared in Fig.~\ref{fig:teff}.
There is very good agreement between the methods
and the deviation is below $1~\sigma$ for most stars.
This indicates that the uncertainties are slightly overestimated.
The spectroscopic and interferometric \teff\ have a mean offset of $-42\pm46$~K 
(rms error; ten stars). This indicates a slight systematic offset
which we round off to $-40\pm20$~K (error on the mean value).
The interferometric and photometric \teff\ have a mean
offset of $-31\pm86$~K (rms error; 6 stars).
Finally, comparing the spectroscopic and photometric data we find a mean difference
of $-24\pm69$~K (rms error; 16 stars). A further examination of the \cite{holmberg07} calibration
was done by \cite{holmberg09}. They compared \teff\ from the photometric calibration 
with interferometric and spectroscopic methods and the infrared flux method. 
They found good agreement with mean offsets no larger than 55~K.

Accurately calibrated \str\ data is extremely useful for faint targets
or large ensembles of stars when it is not feasible to collect a high quality spectrum of each star.
However, interstellar reddening is often a limiting factor on the accuracy.
To estimate the uncertainty on the photometric \teff\ we 
used the calibration error (60~K; \citealt{holmberg07}) and assumed 
errors of 0.005~mag on $b-y$ and 0.07~dex of \feh.
This gives an internal precision of about 70~K. From the comparison with interferometric \teff\ 
we find an rms scatter of about 90~K, which we assign as the accuracy of the photometric \teffs.
To have an idea of the uncertainty when reddening becomes important, we increased 
the error on $b-y$ to 0.015~mag. 
The resulting internal uncertainty on the \teff\ increase from 70~K to 115~K.

We have compared our \teff\ using the direct method with results from the literature for five stars:
\procyon\ (\teff$=6512\pm49$\,K;  \citealt{allende02}),
\betahyi\ (\teff$=5872\pm44$\,K;  \citealt{north07}),
\betavir\ (\teff$=6059\pm49$\,K;  \citealt{north09}),
\etaboo\  (\teff$=6100\pm28$\,K;  \citealt{vanbelle07}), and
\taucet\  (\teff$=5400\pm100$\,K; \citealt{difolco07}).
We find slightly lower values of \teff\ with the mean difference being
$\Delta T_{\rm eff} = -37 \pm 22$\,K (rms error). 


Although we have called the application of Eq.~\ref{eq:teff} a ``direct method'',
in fact \thetald\ depends on the adopted limb-darkening coefficient, which does rely on models.
As mentioned in Sect.~\ref{sec:angular}, 
differences between 1D and 3D models yield a difference of 50~K for \procyon\ \citep{allende02} 
while for \acenb\ it is negligible. 
Another complication is our assumption of zero interstellar reddening, 
which certainly will not hold for more distant stars.
\cite{difolco07} found IR excess around \taucet, which is one of our targets. 
They found the flux in the visible region to be well-fitted by a model with
$T_{\rm eff}=5400\pm100$\,K, which agrees with our our more simple
assumptions resulting in $5383\pm47$\,K. 
We have estimated that the uncertainty on \teff\ from the measurement
errors (\fbol\ and \thetald) is typically $\sigma_{\rm fund}=50$--$60$~K 
(although for \betaaql\ it is $\approx 110$~K).
We estimate that the systematic uncertainty on \teff\ 
(\eg\ from using 1D and not 3D models) is probably about $\sigma_{\rm 1D-3D}=50$~K. 

For our final set of homogeneously determined \teff\ values,
we have adopted the results from 
the spectroscopic analysis of 23 stars (see Sect.~\ref{sec:vwa})
but we apply the determined mean offset of $-40\pm20$~K. 
The offset is valid for stars of spectral type F5--K2 with luminosity class III to V.
The value is rounded is off to 40~K for simplicity and the error on the mean is $\sigma_{\rm spec}=20$~K from ten stars. 
We have added the uncertainties 
as $\sigma^2 = \sigma_{\rm fund}^2 + \sigma_{\rm 1D/3D}^2 + \sigma_{\rm spec}^2
          = (60^2 + 50^2 + 20^2)$\,K$^2$, giving $\sigma = 80$\,K.

The relatively small scatter on the difference between \teff\ from interferometry
and spectroscopy gives us confidence that we can determine the \teff\ to 80~K (1-$\sigma$ uncertainty).
From a high-quality spectrum, this should be possible to achieve for any late-type star.
We mention that the Versatile Wavelength Analysis method (VWA; see Sect.~\ref{sec:atm})
has already been used in the spectroscopic study of several targets of
\corot\ \citep{bruntt09} and \kepler\ \citep{chaplin10}. 
However, in these studies the $-40$~K offset was not known, but we recommend that it is applied.



 \begin{figure*} 
 \begin{center} 
 \includegraphics[width=8.8cm]{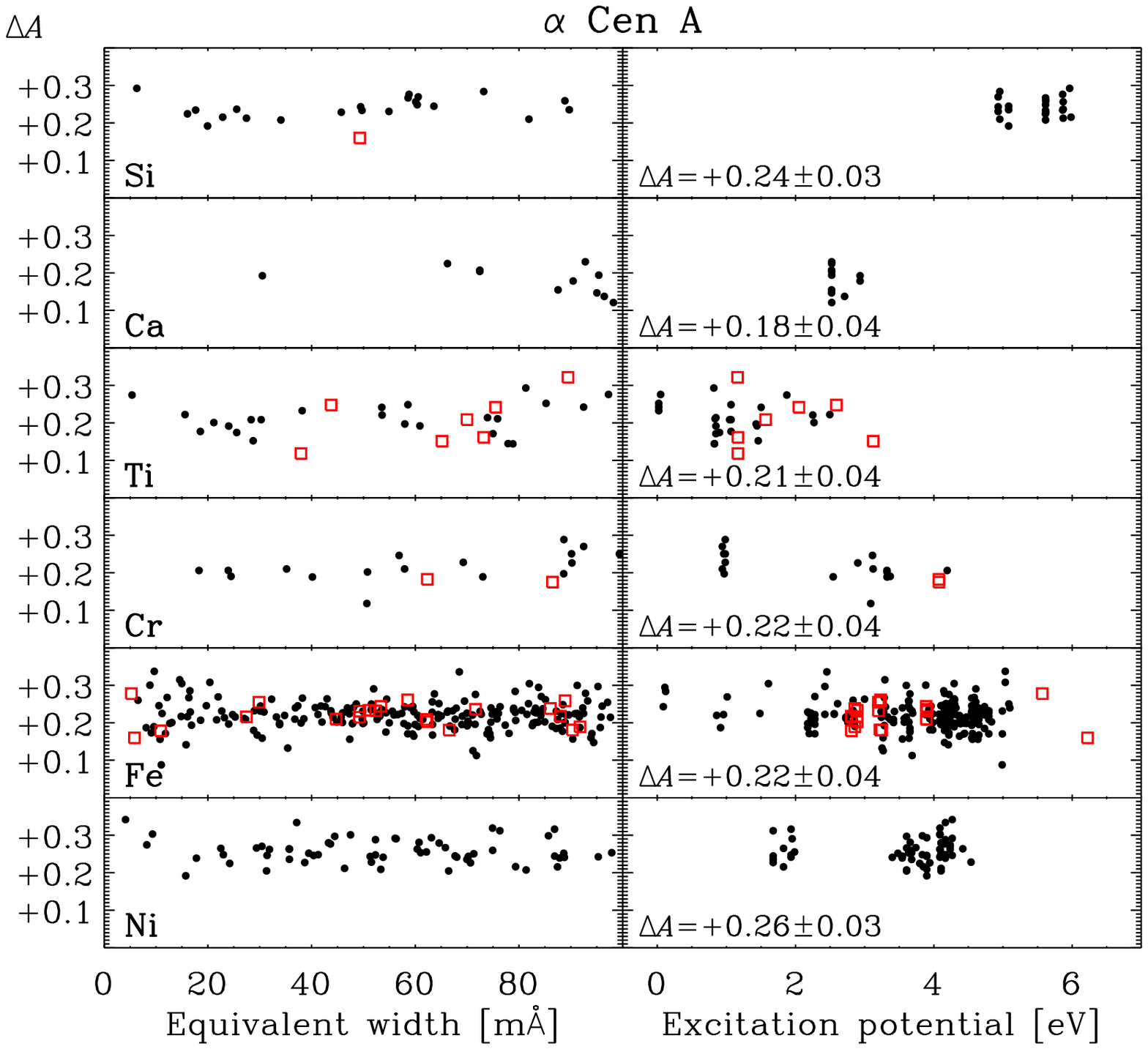}
 \includegraphics[width=8.8cm]{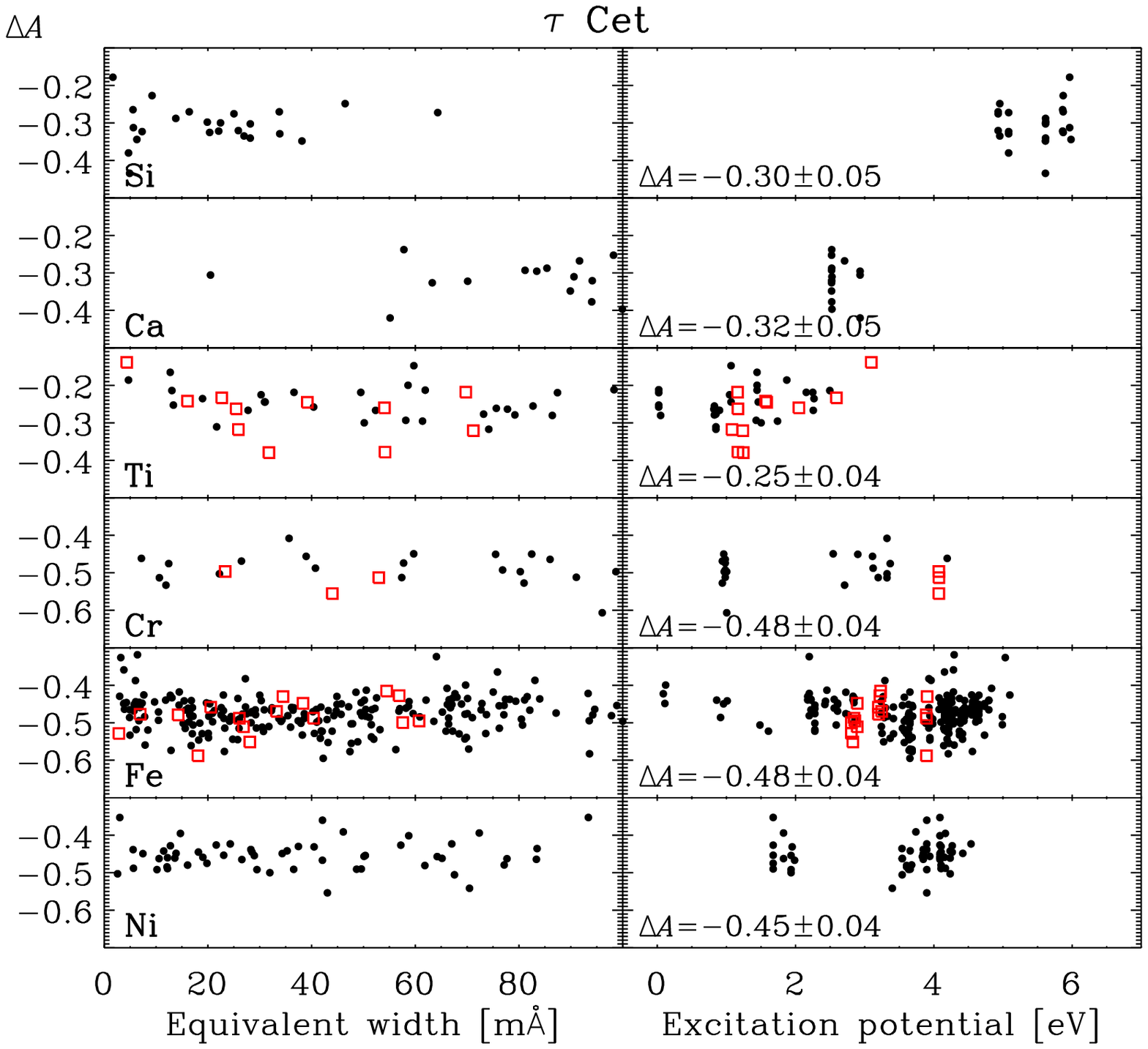}
 \caption{Abundances relative to the Sun for six atomic species in \acena\ and \tauceti.
The abundances are plotted versus equivalent width and excitation potential and no significant correlations are seen,
indicating that the atmospheric model parameters are correct.
The mean abundances and rms scatter determined from the neutral lines are given in the right panels as $\Delta\,A$.
Neutral and singly ionized lines are marked with filled and open symbols, respectively. 
Notice the enhancement of about 0.2~dex for the $\alpha$ elements (Si, Ca, Ti) in the metal-poor star \tauceti.
 \label{fig:abund} }
 \end{center} 
 \end{figure*} 

\section{Spectroscopic analysis}
\label{sec:vwa}

We used high-resolution high-S/N spectra to determine the atmospheric
parameters of the stars: \teff, \logg, chemical composition, and \vsini. 
This is a well-known and widely used ``classical'' analysis method \citep{fuhrmann98, valenti05}, 
but it is indirect since it relies on the adopted model atmospheres.
In this Section, we will determine the chemical composition 
in the photosphere, which provides important input to asteroseismic modelling. 
Also, we will compare the detailed analysis with the metallicity determined from \str\ indices.
Finally, we have analysed non-blended lines to determine 
the line broadening due to \vsini\ and macroturbulence.


\subsection{Spectroscopic observations\label{sec:specobs}}

We used spectra of the 23 targets, primarily from
the High Accuracy Radial velocity Planet Searcher (HARPS spectrograph; $R=115\,000$) 
mounted on the ESO~3.6\,m telescope at La Silla, Chile. 
The exceptions are the spectrum of \betahyi\ from 
the University College of London Echelle Spectrograph (UCLES; $R=65\,000$) at 
the 3.9\,m Anglo-Australian Telescope and the spectrum of \etaboo\ from 
the FIbre-fed Echelle Spectrograph (FIES; $R=68\,000$) at the 2.5\,m Nordic Optical Telescope.
In most cases the spectra have been observed as part of asteroseismic
campaigns, which means several hundred spectra are available in some cases.
We selected typically 20--60 spectra of each target, collected on the
same night, and these were co-added after discarding spectra with poor S/N.
The S/N of the spectra was calculated in continuum windows in the range 5800--6200\,\AA,
with typical values from 400--800. 
In Fig.~\ref{fig:spec} we show examples of spectra for six of the target stars.
Although the shown spectral range contains rich information, 
it comprises only 1\percent\ of the entire available range. 
The wide Ca~line at $\lambda6162$\,\AA\ is one of the indicators used 
to constrain the surface gravity (Sect.~\ref{sec:logg}).

We used pipeline-reduced spectra which in most
cases automatically merges the overlapping echelle orders. 
For this reason we chose not to use the Balmer lines,
although they could be used to constrain \teff\ for the early-type stars in the sample.
The spectra from the pipeline were normalized by
identifying continuum windows in the spectrum. This was done manually
by comparing the observed spectrum with a template spectrum computed
with the same approximate parameters (\teff, \logg, \feh, and \vsini) as 
the observed star. The template was very useful for identifying ``true'' 
continuum windows or regions where the rectification of the spectrum
cannot be done objectively. For a detailed description of the 
rectification process we refer to \fridlund. 

The abundance analysis was based on 1D atmospheric
MARCS models \citep{gustafsson08} that assume local thermodynamic equilibrium (LTE).
We used atmosphere models interpolated in a grid and 
atomic line data are from VALD \citep{kupka99}. 
For the region around lithium at $\lambda6707.8$\,\AA, 
we used atomic data from \cite{ghezzi09},
although we did not include the weak CN bands. 
The model atmospheres used the recently revised 
Solar abundances \citep{grevesse07} and included
$\alpha$-element enhancement for metal-poor stars.
To correct for the short-comings of the 1D~LTE analysis to first order,
we measured abundances differentially with respect to a Solar spectrum (see \citealt{bruntt08}).
However, we caution that some elements will be affected by non-LTE effects,
especially for cool and low-metallicity stars (for a discussion, see \citealt{asplund05}).
Our reference spectrum was taken from the book by \cite{hinkle00} 
(published in original form by \citealt{kurucz84}). 
This spectrum was acquired with the Fourier Transform Spectrometer
(FTS) on the McMath Solar Telescope at Kitt Peak National Observatory.

The design of the FTS ensures that there is no scattered light.
However, our observations were made with three different
echelle spectrographs that could potentially be affected by scattered light.
For the \fies\ spectrum of \etaboo\ we found \feone\ 
abundances to be $\approx 0.1$~dex lower for lines with 
wavelengths shorter than 5500\,\AA.
This could be an indication for the presence of scattered light or improper
subtraction of the flux between the closely spaced echelle orders at short wavelengths.
A similar effect was seen for the \ucles\ spectrum of \betahyi. 
For this reason we only used spectral lines above 5500\,\AA\ for \etaboo\ and \betahyi.
The effect was not seen in the \harps\ spectra and so the entire spectral range was used.



\subsection{\mbox{\boldmath \teff}, \mbox{\boldmath \logg}, microturbulence from iron lines\label{sec:atm}}

We used the \vwa\ software \citep{bruntt04, bruntt08, bruntt09} to analyse the spectra.
This semi-automatic software package selects the 
least blended lines and iteratively fits the abundances.
Lines were selected for each star and the fit of the synthetic spectrum was carefully
investigated. Typically, $600$--$1000$ lines could be used, 
although for the relatively fast rotators (\taupsa, \etaboo\ and \gammaser) we used $250$--$400$ lines. 
The part of spectrum with most of the lines ranged from $4880$ to $6820$\,\AA, while
we generally avoided regions affected by telluric lines 
(\eg\, $5880$--$6000$ and $6275$--$6330$, and $6460$--$6600$\,\AA).
We initially used a model atmosphere with \logg\ and \teff\ as determined from the spectral type
and solar metallicity was assumed. 
The atmospheric parameters and the microturbulence were then refined in several steps.
This was done by minimizing the correlations between 
the abundance of Fe\,{\sc i} lines and equivalent width and excitation potential,
and requiring good agreement between the abundances of the neutral and ionized species of Fe, Cr and Ti. 

As examples of the quality of the data, we show the abundance of 
Si, Ca, Ti, Cr, Fe and Ni for \acena\ and \taucet\ in Fig.~\ref{fig:abund}. 
The mean abundance and rms scatter are indicated in the right panels for the neutral lines.
It is seen that the metal-poor star \taucet\ has 
a relatively high abundance of the $\alpha$-elements (Si, Ca, and Ti) compared to the other metals.

One of the best-studied solar-type stars is \acena\ and 
\citealt{mello08} provided a list of 17 studies.
Among the 13 that were based on spectroscopic data there is 
very good agreement on \teff\ and \logg, and most studies indicate
${\rm [Fe/H]} \approx +0.25$. The two most detailed studies of \acena\ 
are those of \cite{neuforge97} and \cite{mello08} who adopted 
a differential analysis similar to ours.
Their results for the spectroscopic parameters agree with ours, 
but we include 3--6 times as many spectral lines. 
While this allows us to derive the atmospheric parameters with lower intrinsic error, 
it is the systematic errors that dominate the uncertainty.

 \begin{figure} 
 \begin{center} 
 \includegraphics[width=8.8cm]{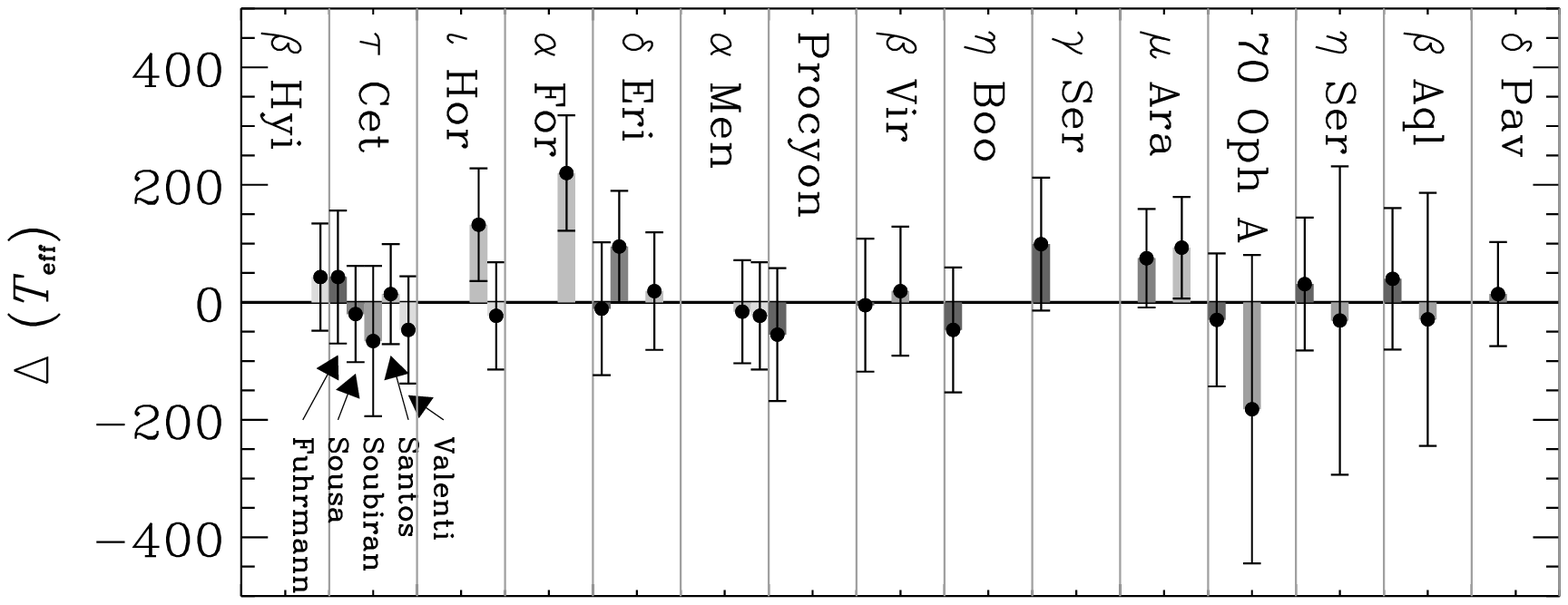}
 \includegraphics[width=8.8cm]{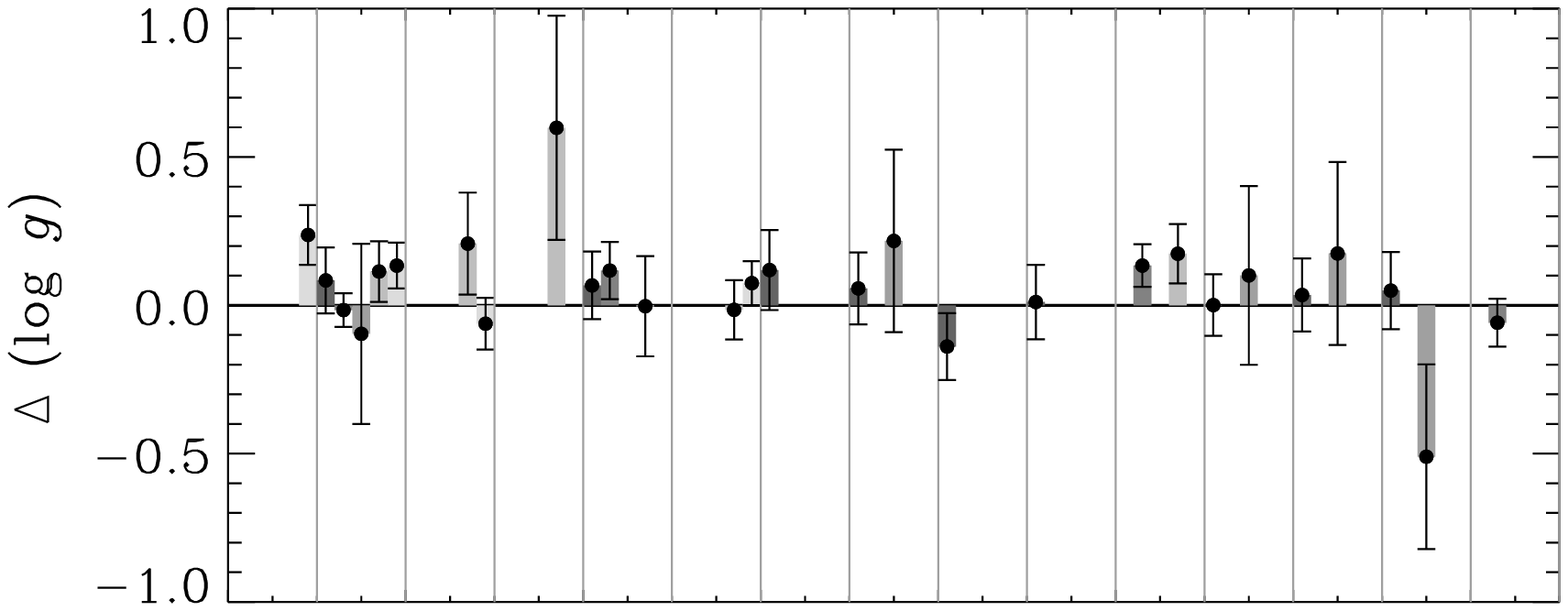}
 \includegraphics[width=8.8cm]{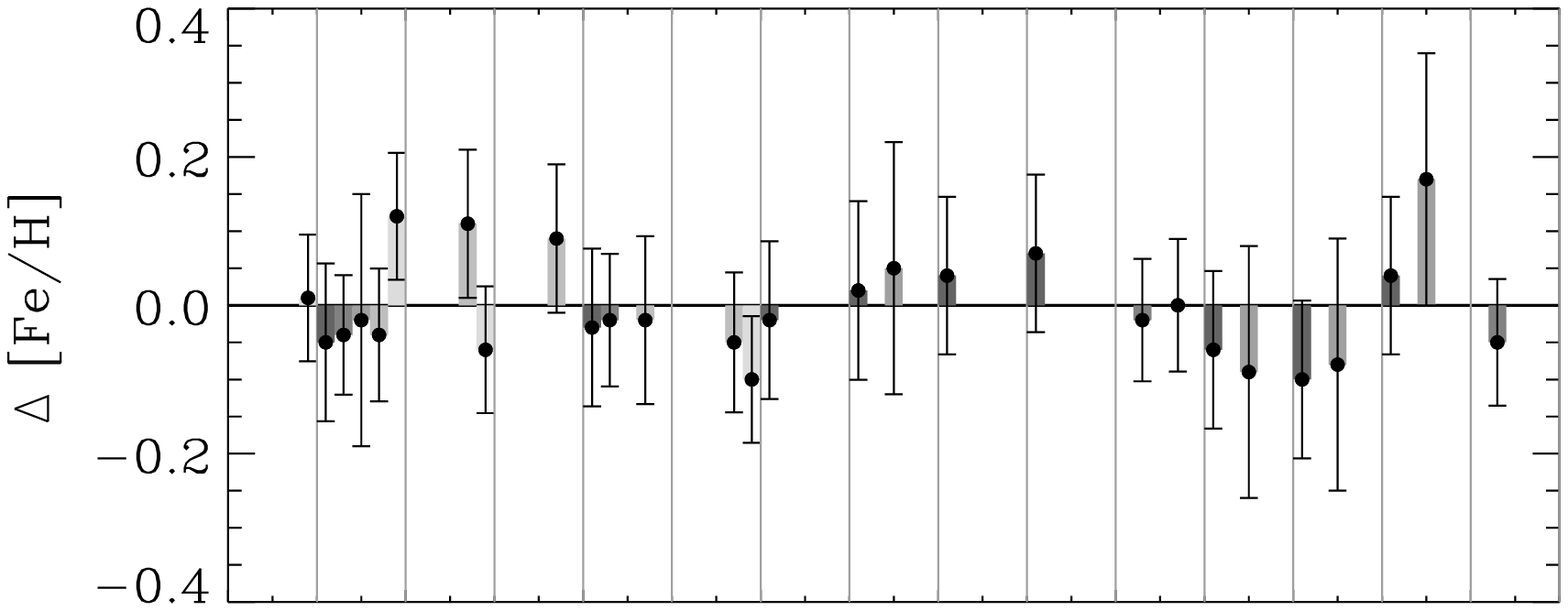}
 \caption{Atmospheric parameters from five studies in the literature 
compared to values from \vwa. 
Individual value are given in Appendix~\ref{app:tables}.
 \label{fig:atmos} }
 \end{center} 
 \end{figure} 

Several of the targets have been analysed using similar techniques in the literature.
As a consistency check we have compared the atmospheric parameters with 
similar recent studies that also use 1D~LTE models and similar recipes for
adjusting \teff, \logg\ and microturbulence. In Fig.~\ref{fig:atmos} we compare
our \teff, \logg\ and \feh\ with five studies indicated by the gray bars. From left
to right it is \cite{fuhrmann98, fuhrmann04, fuhrmann08},
\cite{sousa08}, \cite{soubiran98}, \cite{santos04, santos05}, and \cite{valenti05}. 
In Appendix~\ref{app:tables} we list the values from the literature.
The maximum deviations are 220~K in \teff, 0.58 dex in \logg, and 0.24 dex in \feh.
These are quite large deviations compared to the typical uncertainties we have determined.
However, for the most deviant parameters, the literature nearly always quote large uncertainties.
For the \logg\ of \alphafor, \cite{santos04} found $4.40\pm0.37$ while we have 
                                                   $3.82\pm0.08$. 
The high \logg\ value explains the high \teff\ found by \cite{santos04} ($\Delta T_{\rm eff}=220$~K),
since \logg\ and \teff\ are not independent.
For the \logg\ of \betaaql,  \cite{soubiran98} found $3.04\pm0.30$ where we get 
                                                     $3.61\pm0.08$ in agreement with \cite{fuhrmann04} who found 
                                                     $3.60\pm0.10$. 
For the same star, \cite{soubiran98} found ${\rm [Fe/H]} = -0.04\pm0.15$, 
                                            while we find $-0.20\pm0.07$, which is in agreement with \cite{fuhrmann04} who found 
                                                          $-0.17\pm0.07$.

We conclude that our spectroscopic analysis is in good agreement with previous studies.
In several cases we provide more accurate parameters, which was possible 
due to the high quality of the spectra. 
The most important aspect of our results is that 
parameters for all 23 stars were determined in a homogeneous way.

\subsection{Surface gravity from pressure-sensitive lines\label{sec:logg}}

For cool stars there are strong pressure-sensitive lines
that can be used to constrain the \logg\ parameter.
Commonly used lines are \mgone, \naone, and the calcium lines at $\lambda6122$\,\AA\
and $\lambda6262$\,\AA\ \citep{gray05}.
We follow the approach described by \cite{fuhrmann97} in 
which the Van der Waals constants are adjusted until the
\logg\ parameter found for a spectrum of the Sun is consistent with the canonical
value, $\log g_\odot = \log {\rm M}_\odot / {\rm R}_\odot^2 = 4.437$. We used the FTS spectrum
of the Sun and checked that a solar spectrum of the Sun from \harps\ gave consistent results
(for details see \fridlundalt).
When fitting the wide lines we first determine the abundance of the element from
weaker lines. The same is true for \vsini\ and \vmacro\ although the result is less 
sensitive to this. 

\begin{figure} 
\begin{center} 
\includegraphics[width=9cm]{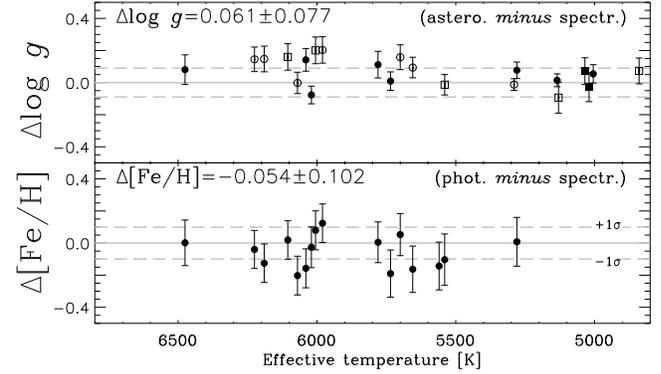}
\caption{
Top panel: Difference in \logg\ determined from asteroseismic information and spectroscopic analysis. 
Solid symbols are used for stars with measured angular diameters.
Bottom panel: Difference in metallicity determined from \str\ indices and spectroscopic determination.
The individual values are given in Table~\ref{tab:fund} and \ref{tab:logg}.
\label{fig:logg} }
\end{center} 
\end{figure} 

A serious problem, especially for early-type stars is the normalization
of the continuum around the \mgone\ lines around 5167, 5173 and 5184\,\AA. 
The lines are so wide that it is difficult to normalize the spectra objectively,
and two of the lines lie close and therefore have no continuum between them.
We found that the continuum in the red wing (around 5160\,\AA)
is impossible to define for stars earlier than K0, 
and therefore only the 5184\,\AA\ line can be used reliably.
We only used the wings of the \naone\ lines to check for consistency, 
since the region is affected by telluric lines. 
The Ca lines are not as wide as the \mgone\ lines and it is easier 
to define the continuum. Another advantage of the Ca lines is that there are
several weak lines available, so the Ca abundance is well-determined. 
There are only a few weak Mg lines available, 
and this affects the accuracy of \logg\ determined from the \mgone\ lines.

%
%
 \begin{figure*} 
 \begin{center} 
 \includegraphics[angle=90,width=18cm]{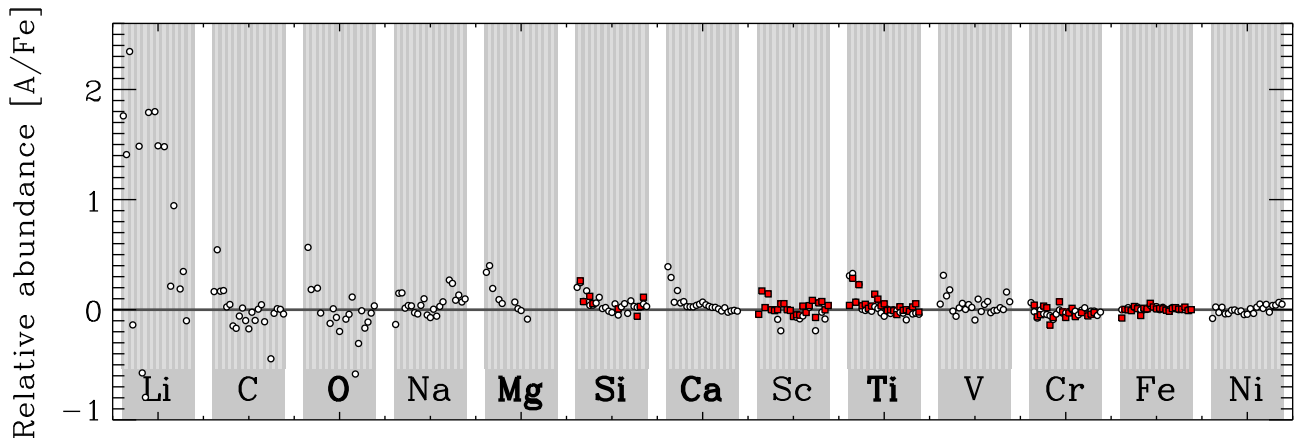}
 \caption{Mean abundances of 13 elements in 23 solar-type stars sorted by
\feh\ increasing from left to right. To be able to show the results on 
the same scale abundances are measured relative to \feone.
The mean abundance from neutral and ionized lines are marked with circles and box symbols, respectively. 
The names of the $\alpha$ elements are written with bold font:
notice the generally higher abundance for the low metallicity stars.
Abundances for each individual star is given in Table~\ref{tab:ab}.
 \label{fig:pattern} }
 \end{center} 
 \end{figure*} 

To fit the observed spectrum we calculated synthetic spectra for three values of \logg. 
We then calculated the \chisq\ value for a few selected regions that are not seriously
affected by blending lines. \fridlund\ gave a more 
detailed description of the method when applied to \acenb.
In Table~\ref{tab:logg} we compare the \logg\ values 
determined from different techniques. It is seen that the precision on the
\logg\ determination changes a lot from star to star and is due to the
change in sensitivity depending on the spectral type. 
In general, the Ca lines at $\lambda6122$\,\AA\ and $\lambda6262$\,\AA\ are the 
most useful, but the \mgone\ line is applicable for the early-type stars.
For the final spectroscopic value of \logg\ we 
calculated the weighted mean value which is given in column 4.

In Fig.~\ref{fig:logg} (top panel) we compare the \logg\ values determined from spectroscopy
(weighted mean value using \feonetwo\ and the Ca lines)
with that from the combination of radius and mass through 
\begin{equation}
\log g = \log M / {\rm M}_\odot - 2\, \log R / {\rm R}_\odot + \log g_\odot.
\label{eq:logg}
\end{equation} 
Different symbols in Fig.~\ref{fig:logg} are used to indicate the source of $M$ and $R$.
The radius is from Sect.~\ref{sec:radius}, as found using either 
interferometric measurements (filled symbols) or the combination of 
luminosity and effective temperature (open symbols). 
The mass was determined from the asteroseismic information (Sect.~\ref{sec:mass}) 
using either the \largesep\ (circles) or \numax\ (box symbols).
There is generally good agreement, with a mean difference of $+0.061\pm0.077$ dex.
Instead of using the intrinsic uncertainties on \logg\ given in Table~\ref{tab:logg},
we shall adopt 0.08~dex (the rms scatter) as the uncertainty on \logg\ for all 
stars in the sample. This is very similar to the uncertainty adopted 
in other studies of large samples of stars \citep{valenti05, fuhrmann04}.

Finally, we find excellent agreement ($<0.1$\,dex) when comparing 
the results in Table~\ref{tab:logg} for the four binary stars,
for which absolute $M$ and $R$ is available (although $R$ for \opha\ is not a direct measurement).
This further supports that the indirect methods give 
consistent results for the surface gravity.




%
%
\begin{table*}
 \centering
  \caption{Surface gravities from methods (columns 2--9).
The \vmacro, \vmicro\ and \vsini\ from the spectroscopic analysis (columns 10--12) 
have uncertainties of 0.6, 0.07, and 0.6\,\kms.
\label{tab:logg}}
 \setlength{\tabcolsep}{2pt} 
   \begin{tabular}{lcc|cccccc|ccr}
   \hline
       & Binary &   Asteroseis. & Mean of        & \multicolumn{5}{c}{Spectroscopic methods} & \vmacrooo & \vmicrooo & \vsini \\
Star &  $M+R$ & $\rho,R$      & spec.\ methods & \feonetwo & \mgone & Ca-6122 & Ca-6162 & Ca-6439 & \kmss & \kmss & \kmss \\
\hline

 \betahyi & $             $& $3.955\pm0.018$&$3.843\pm0.081$&$ 3.92\pm 0.11$&$ 3.68\pm 0.32$&$ 3.72\pm 0.18$&$ 3.81\pm 0.18$&$ 3.86\pm 0.81$ & $ 2.9$ &$ 1.16$ & $ 2.7$ \\ 
 \taucet  & $             $& $4.533\pm0.018$&$4.456\pm0.048$&$ 4.48\pm 0.10$&$ 4.91\pm 0.22$&$ 4.39\pm 0.10$&$ 4.42\pm 0.07$&$ 4.57\pm 0.24$ & $ 0.9$ &$ 0.72$ & $ 0.7$ \\
 \iotahor & $             $& $4.399\pm0.022$&$4.402\pm0.063$&$ 4.41\pm 0.11$&$ 4.37\pm 0.12$&$ 4.41\pm 0.16$&$ 4.39\pm 0.13$&$ 4.70\pm 0.42$ & $ 3.3$ &$ 1.04$ & $ 5.3$ \\
 \alphafor& $             $& $4.003\pm0.033$&$3.802\pm0.077$&$ 3.82\pm 0.11$&$ 3.72\pm 0.13$&$ 3.85\pm 0.21$&$ 4.08\pm 0.34$&$ 4.22\pm 0.89$ & $ 3.7$ &$ 1.29$ & $ 3.9$ \\
 \deltaeri& $             $& $3.827\pm0.018$&$3.773\pm0.054$&$ 3.74\pm 0.10$&$ 3.64\pm 0.39$&$ 3.74\pm 0.11$&$ 3.85\pm 0.09$&$ 3.69\pm 0.19$ & $ 0.9$ &$ 0.89$ & $ 0.7$ \\
 \alphamen& $             $& $             $&$4.425\pm0.044$&$ 4.45\pm 0.10$&$ 4.43\pm 0.12$&$ 4.36\pm 0.13$&$ 4.42\pm 0.06$&$ 4.55\pm 0.25$ & $ 1.0$ &$ 0.93$ & $ 0.6$ \\
 \procyon & $3.976\pm0.016$& $3.972\pm0.018$&$3.891\pm0.090$&$ 4.01\pm 0.14$&$ 3.80\pm 0.12$&$ 3.86\pm 0.75$&$ 3.90\pm 0.43$&$ 4.46\pm 0.70$ & $ 4.6$ &$ 1.69$ & $ 2.8$ \\
 \pup     & $             $& $4.244\pm0.023$&$4.088\pm0.074$&$ 4.17\pm 0.10$&$ 3.98\pm 0.12$&$ 4.37\pm 0.58$&$ 4.03\pm 0.32$&$ 4.26\pm 0.89$ & $ 1.9$ &$ 1.21$ & $ 1.6$ \\
 \ksihya  & $             $& $2.883\pm0.032$&$2.809\pm0.077$&$ 2.76\pm 0.10$&$ 2.37\pm 0.42$&$ 2.85\pm 0.18$&$ 3.10\pm 0.20$&$ 2.65\pm 0.39$ & $ 3.8$ &$ 1.20$ & $ 2.4$ \\
 \betavir & $             $& $4.125\pm0.018$&$3.983\pm0.068$&$ 4.01\pm 0.11$&$ 4.01\pm 0.16$&$ 3.91\pm 0.12$&$ 4.06\pm 0.21$&$ 4.04\pm 0.54$ & $ 3.6$ &$ 1.29$ & $ 2.0$ \\
 \etaboo  & $             $& $3.822\pm0.019$&$3.899\pm0.052$&$ 3.88\pm 0.11$&$ 3.87\pm 0.06$&$ 4.07\pm 0.27$&$ 4.08\pm 0.19$&$ 4.20\pm 0.49$ & $ 5.3$ &$ 1.74$ & $11.9$ \\ 
 \acena   & $4.307\pm0.005$& $4.318\pm0.017$&$4.309\pm0.055$&$ 4.28\pm 0.10$&$ 4.32\pm 0.11$&$ 4.28\pm 0.21$&$ 4.32\pm 0.09$&$ 4.60\pm 0.43$ & $ 2.3$ &$ 1.00$ & $ 1.9$ \\
 \acenb   & $4.538\pm0.008$& $4.530\pm0.018$&$4.515\pm0.036$&$ 4.45\pm 0.10$&$ 4.01\pm 0.50$&$ 4.53\pm 0.07$&$ 4.51\pm 0.05$&$ 4.65\pm 0.16$ & $ 0.8$ &$ 0.83$ & $ 1.0$ \\
\hdcarrier& $             $& $4.229\pm0.023$&$4.083\pm0.077$&$ 4.05\pm 0.14$&$ 4.02\pm 0.14$&$ 4.15\pm 0.13$&$ 4.17\pm 0.28$&$ 4.22\pm 0.89$ & $ 4.1$ &$ 1.31$ & $ 5.2$ \\
\gammaser & $             $& $4.169\pm0.032$&$4.009\pm0.076$&$ 4.01\pm 0.11$&$ 3.93\pm 0.11$&$ 4.33\pm 0.46$&$ 4.48\pm 0.32$&$ 5.33\pm 1.17$ & $ 3.9$ &$ 1.29$ & $10.2$ \\
 \muara   & $             $& $4.228\pm0.023$&$4.136\pm0.060$&$ 4.25\pm 0.10$&$ 4.04\pm 0.16$&$ 3.92\pm 0.20$&$ 4.12\pm 0.10$&$ 4.03\pm 0.39$ & $ 2.6$ &$ 1.05$ & $ 1.4$ \\
 \opha    & $4.468\pm0.030$& $4.555\pm0.023$&$4.569\pm0.028$&$ 4.59\pm 0.10$&$ 4.50\pm 0.23$&$ 4.53\pm 0.06$&$ 4.57\pm 0.04$&$ 4.73\pm 0.15$ & $ 1.5$ &$ 0.93$ & $ 0.9$ \\
 \etaser  & $             $& $3.029\pm0.037$&$2.955\pm0.072$&$ 2.96\pm 0.10$&$ 2.90\pm 0.25$&$ 2.88\pm 0.25$&$ 2.98\pm 0.13$&$ 3.03\pm 0.55$ & $ 2.1$ &$ 1.04$ & $ 1.1$ \\
 \betaaql & $             $& $3.525\pm0.036$&$3.550\pm0.083$&$ 3.49\pm 0.10$&$ 3.49\pm 0.33$&$ 3.67\pm 0.37$&$ 3.76\pm 0.19$&$ 3.65\pm 0.52$ & $ 2.1$ &$ 0.88$ & $ 0.9$ \\
 \deltapav& $             $& $4.306\pm0.034$&$4.319\pm0.054$&$ 4.30\pm 0.10$&$ 4.43\pm 0.26$&$ 4.16\pm 0.15$&$ 4.34\pm 0.08$&$ 4.48\pm 0.22$ & $ 1.7$ &$ 0.95$ & $ 1.0$ \\
 \gammapav& $             $& $4.397\pm0.022$&$4.195\pm0.079$&$ 4.24\pm 0.11$&$ 4.06\pm 0.14$&$ 4.33\pm 0.27$&$ 4.26\pm 0.24$&$ 4.39\pm 1.48$ & $ 2.0$ &$ 1.19$ & $ 1.8$ \\
 \taupsa  & $             $& $4.240\pm0.021$&$4.096\pm0.073$&$ 4.09\pm 0.14$&$ 4.05\pm 0.10$&$ 4.27\pm 0.25$&$ 4.24\pm 0.26$&$ 5.08\pm 0.71$ & $ 4.1$ &$ 1.30$ & $13.6$ \\
 \nuind   & $             $& $3.432\pm0.035$&$3.526\pm0.090$&$ 3.48\pm 0.10$&$ 4.50\pm 0.38$&$ 3.38\pm 0.28$&$ 3.49\pm 0.47$&$ 2.84\pm 1.19$ & $ 1.4$ &$ 1.35$ & $ 1.1$ \\

\hline
\end{tabular}
\end{table*}

\subsection{Chemical composition\label{sec:chem}}

With the atmospheric parameters determined
from \feonetwo\ and the pressure-sensitives lines,
we computed the mean abundances for 13 elements. 
In Table~\ref{tab:ab} we list the mean abundances and in Fig.~\ref{fig:pattern}
we plot the abundances. To be able to show the results on the same
scale we have offset the abundances by the mean abundance of \feone\ lines.
It can be seen that \feonetwo\ agree since this was a requirement 
when adjusting the atmospheric parameters. There is some
scatter in the light elements (Li, C, O), 
while most of the metals have quite low scatter. 
This indicates that a simple scaling from the solar abundance using just \feh\ is
a good approximation. However, for the stars with low metallicity we see a clear
increase in the abundance of the $\alpha$ elements (Ca, Si and Ti). 
This $\alpha$ enhancement is shown in more detail for \taucet\ in Fig.~\ref{fig:abund}.

\cite{holmberg07} presented a calibration of \feh\ based on \str\ indices and
in the last two columns in Table~\ref{tab:fund} we compare them 
with the spectroscopic values. The comparison is shown in Fig.~\ref{fig:logg} in the
bottom panel. There is excellent agreement with a mean difference of
$\Delta {\rm [Fe/H]} = -0.054\pm0.102$ (rms scatter).

\subsection{Projected rotational velocity and macroturbulence
\label{sec:vsini}}

The mean rotational velocity is important for asteroseismic modelling since the
observed frequencies are split depending on the projected rotation rate.
The high-precision photometry from \corot\ and \kepler\ provides indirect measures
of the rotational period for solar-type star as spots traverse the surface. 
However, in the case of the solar-type \corot\ target HD~49385, 
Deheuvels~et~al. (A\&A, 2010, accepted) could not directly measure the 
rotation period from the photometry. Instead they used the spectroscopically 
measured \vsini\ to rule out one of the possible scenarios when interpreting 
the asteroseismic data.


The shape of a spectral line depends on various 
physical processes on a microscopic scale 
(atomic absorption, pressure and thermal broadening)
but also depends on the macroscopic velocity fields in the photosphere
due to convection cells and the rotation of the star. 
For some lines the detailed line shape is further affected by 
hyperfine structure and the Zeeman effect in the case of strong magnetic fields.


We used isolated spectral lines to determine
the projected rotational velocity (\vsini) and macroturbulence
of the target stars following similar assumptions and methods
as \cite{saar97} and \cite{reddy02}. 
In our analysis we described the broadening mechanisms with some common simplifications.
We included the broadening due to the projected rotational
velocity (\vsini) in the formulation of \cite{gray05},
instrumental broadening (a Gaussian profile),
macroturbulence (also a Gaussian profile),
and line broadening parameters from the \vald\ database. 
In addition, when calculating the synthetic profiles
we introduced the microturbulence parameter. 
We note that from bisector analysis of the asymmetry of spectral lines much
more can be learned about the granulation. This requires individual 
spectra with very high resolution and is beyond 
the scope of the current work \citep{dravins08, allende02}.


Since we have used different spectrographs, 
we made careful estimations of the instrumental profiles. 
This was done by measuring the widths
of several telluric lines from 6279--6304\,\AA. 
Up to 19 (typically 12--15) lines could be used depending on how 
the telluric lines were shifted and blended with the stellar lines. 
The telluric lines were fitted with Gaussian profiles and 
the instrumental resolution was estimated from the mean FWHM ($\Delta \lambda$) 
of these profiles,  \ie\ $R=\langle \lambda / \Delta \lambda \rangle$.
In general we measured a resolution 10--15\% lower than the nominal
values found in the instrument descriptions. 
We have assumed that the telluric lines have zero width, but in fact telluric lines 
have non-zero widths that change with observing conditions and airmass.
Thus, our assumption will indeed give a tendency to 
underestimate the resolution of the spectrograph.
For the HARPS spectra we also measured $R$ from the Th-Ar calibration spectra,
and in this case we measured $R=107\,000\pm3000$ (rms scatter from eight lines), which
is close to the instrumental specification ($R=115\,000$, \citealt{mayor03}).



To select the most suitable lines for the determination of \vsini\ we chose
lines from the abundance analysis that are present in at least 60\% of all stars.
All lines are relatively weak, with equivalent widths in the Solar spectrum
between 20 and 100\,m\AA. For each line we fixed the fitted values of the abundance
and the instrumental resolution. We then convolved the synthetic
line by different combinations of \vsini\ and \vmacro\ in a regular grid with
steps of 0.15\,\kms. We calculated the $\chi^2$ value of each fit from the sum
over a region around each 
line\footnote{The range is typically $\pm0.2$ to $\pm0.5$\,\AA\ for \vsini\ values from 1 to 15\,\kms.}:
\begin{equation}
\chi^2 = {1 \over n} \displaystyle\sum_{i=1}^N { (O_i - S_i)^2 \over \sigma^2 },
\end{equation}
where $N$ is the total number of observed data points and
$n$ is the number of degrees of freedom in the fit: $n=N - 3$, since
three parameters are being fitted: wavelength shift, \vsini\ and \vmacro.
$O_i$ and $S_i$ denote the relative flux of the observed and synthetic profile
and $\sigma$ is the measured noise in the continuum (see \citealt{reddy02}).
Typically 10--30 lines are used for each star. The final values of \vsini\
and \vmacro\ are calculated as the mean value and the rms value is taken as
the 1-$\sigma$ uncertainty.

 \begin{figure} 
 \begin{center} 
 \includegraphics[width=8.9cm]{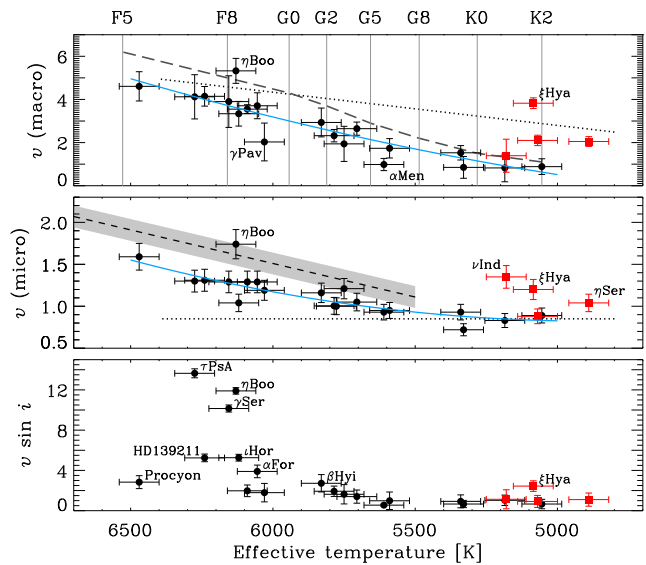}
 \caption{Macro- and microturbulence and \vsini\ determined from the spectroscopic analysis.
In the two top panels the solid line is a second order fit to the data (Eqs.~\ref{eq:calone} and \ref{eq:caltwo}).
In the top panel the dashed line is adopted from Gray~(2005) and in the middle panel
the dashed line and shaded region is the calibration of \vmicro\ from Edvardsson et al.~(1993).
The dotted lines are the results from Valenti \& Fisher (2005) (see discussion in the text).
 \label{fig:vsini} }
 \end{center} 
 \end{figure} 

In the bottom panel in Fig.~\ref{fig:vsini} we show the \vsini\ values of the target stars,
showing that there are more fast rotators among the early type stars.
We have compared the determined \vsini\ and \vmacro\ 
with similar studies in the literature in Table~\ref{tab:vsini} in Appendix~\ref{ap:vsini}.
We find a good agreement with \cite{saar97} for four stars.
We have six stars in common with \cite{valenti05}, but they did not include
\vmacro\ in their description of the broadening. We find that by quadratically 
adding our values for \vsini\ and \vmacro\ we get good agreement with the \vsini\ values from \cite{valenti05}.
\cite{dravins93} and \cite{allende02} used high-resolution spectra of \betahyi\ and \procyon. 
They used 3D time-dependent atmosphere models but we still
get almost perfect agreement with these two studies for \vsini.



\subsection{Calibrations of macro- and microturbulence\label{sec:turb}}

In Fig.~\ref{fig:vsini} we show the determined values of \vmacroo\ and \vmicroo\
(also given in Table~\ref{tab:logg}). In the top panel we have marked 
the approximate spectral type following the tabulation by \cite{gray05}.
The solid lines are polynomial fits to the data 
while discarding the four giant stars and two apparent outliers 
(\etaboo\ and \alfmen). The fits are expressed as a function of \teff:
\begin{equation}
v_{\rm macro} = 2.26 + 2.90 \, 10^{-3} \Delta T + 5.86 \, 10^{-7} \Delta T^2 \,  {\rm km\,s}^{-1}
\label{eq:calone}
\end{equation}
\begin{equation}
v_{\rm micro} = 1.01 + 4.56 \, 10^{-4} \Delta T + 2.75 \, 10^{-7} \Delta T^2 \, {\rm km\,s}^{-1},
\label{eq:caltwo}
\end{equation}
where $\Delta T = T_{\rm eff}^\star - 5700$\,K; 
$T_{\rm eff}^\star$ is the spectroscopic value.
These calibrations are valid for unevolved stars ($\log g > 4.0$) in the range 5000--6500\,K and 
the rms scatter for the fits are 0.27 and 0.09 \kms, respectively.

The dashed line in the top panel in Fig.~\ref{fig:vsini} is from \cite{gray05} 
(see his Fig.~17.10 for luminosity class~V) and the dotted line is the
upper limit on \vmacro\ from \cite{valenti05}.
We find a systematically lower \vmacroo\ for all stars. 
However, the source of the data in \cite{gray05} is not known, 
so we cannot identify the reason for the apparent discrepancy.
To determine upper limits on \vmacro, \cite{valenti05} assumed $v \sin i = 0$\,\kms,
so it is not surprising that we find lower values for all spectral types.
In the middle panel of Fig.~\ref{fig:vsini} we compare our \vmicroo\ values with the 
calibration of \cite{edvardsson93}, valid in the range from 5500--6800~K:
$v_{\rm micro} = 1.25 + 8 \times 10^{-4} \,  (T_{\rm eff} - 6000) - 1.30 \, (\log g - 4.5)$.
The dashed line corresponds to $\log g = 4.3$ and the gray shaded area 
marks the range for changes in \logg\ of $\pm0.1$~dex.
\cite{edvardsson93} report that the scatter of individual stars is 0.3~\kms\ around this relation.
They used only a very limited number lines of \feone\ and \nione\ (up to 17) to constrain \vmicroo,
while we have used hundreds of lines in each star.
However, the larger sample of \cite{edvardsson93} (157 stars) allowed them to
include the variation with \logg. Our \vmicroo\ calibration predicts values that are typically $0.3$~\kms\ lower.
\cite{edvardsson93} used $v_{\rm micro} = 1.15$\,\kms\ whereas we find $0.95$\,\kms\ for the Solar spectrum. 
\cite{valenti05} studied a large sample of over 1000 stars by making 
a ``global fit'' to the observed spectra. They found no dependence of \vmicro\ and \teff\
and adopted a fixed value of 0.85~\kms, indicated by the dotted 
horizontal line in the middle panel in Fig.~\ref{fig:vsini}.
By studying saturated line profiles, \cite{landstreet09} 
demonstrated that higher convective velocity implies a higher microturbulence in A and F type stars.
It is important to note that \vmicro\ is 
a fitting parameter that is only conceptually linked to convective motion on small scales
(see discussions in \citealt{valenti05}, \citealt{gray05}, and \citealt{dravins90-paper4}). 
It is therefore difficult to compare \vmicroo\ for 
different analysis methodologies and different model atmospheres.






%
%
%
\begin{table*}
 \centering
 \caption{Abundances and number of lines used in spectral analysis of 23 target stars, sorted by metallicity.
The uncertainty on the abundances is 0.07 dex. The data are shown graphically in Fig.~\ref{fig:pattern}. 
 \label{tab:ab}}
 \setlength{\tabcolsep}{1.6pt} 
 \begin{footnotesize}
\begin{tabular}{l|lr|lr|lr|lr|lr|lr|lr|lr |lr|lr|lr|lr}


   & \multicolumn{2}{c|}{\nuind} & \multicolumn{2}{c|}{\pup} & \multicolumn{2}{c|}{\gammapav} & \multicolumn{2}{c|}{\taucet} & \multicolumn{2}{c|}{\alphafor} & \multicolumn{2}{c|}{\gammaser} & \multicolumn{2}{c|}{\betaaql} & \multicolumn{2}{c|}{\etaser}
& \multicolumn{2}{c|}{\betahyi} & \multicolumn{2}{c|}{\procyon} & \multicolumn{2}{c|}{\hdcarrier}  & \multicolumn{2}{c|}{\taupsa}\\
\hline

{Li \sc   i} &  $ +0.21   $ &   1 &  $ +0.56  $ &   1 &  $ +1.61   $ &   1 &  $ -0.61   $ &   1 &              &     &  $ +1.22   $ &   1 &  $ -0.73   $ &   1 &  $ -0.91  $ &   1   & $ +1.68   $  &   1 &              &     &  $ +1.74   $ &   1 & $ +1.50   $ &   1 \\ 
{C  \sc   i} &  $ -1.39   $ &   2 &  $ -0.31  $ &   2 &  $ -0.57   $ &   2 &  $ -0.30   $ &   1 &  $ -0.28   $ &   4 &  $ -0.22   $ &   4 &  $ -0.30   $ &   1 &  $ -0.29  $ &   1   & $ -0.16   $  &   1 &  $ -0.04   $ &   4 &  $ -0.15   $ &   5 & $ -0.17   $ &   3 \\ 
{O  \sc   i} &              &     &  $ -0.29  $ &   2 &  $ -0.55   $ &   3 &  $         $ &     &  $ -0.11   $ &   2 &  $ -0.29   $ &   2 &  $         $ &     &  $        $ &       & $ -0.23   $  &   2 &  $ -0.04   $ &   4 &  $ -0.13   $ &   2 & $ -0.19   $ &   2 \\ 
{Na \sc   i} &  $ -1.69   $ &   1 &  $ -0.70  $ &   3 &  $ -0.60   $ &   5 &  $ -0.46   $ &   3 &  $ -0.27   $ &   2 &  $ -0.23   $ &   3 &  $ -0.18   $ &   2 &  $ -0.16  $ &   2   & $ -0.07   $  &   1 &  $ +0.03   $ &   4 &  $ -0.10   $ &   3 & $ -0.06   $ &   2 \\ 
{Mg \sc   i} &  $ -1.21   $ &   1 &  $ -0.45  $ &   2 &  $ -0.54   $ &   2 &              &     &  $ -0.21   $ &   2 &  $ -0.21   $ &   2 &              &     &             &       &              &     &  $ +0.02   $ &   2 &  $ -0.05   $ &   2 & $ -0.00   $ &   2 \\  
{Si \sc   i} &  $ -1.35   $ &   1 &  $ -0.61  $ &  18 &  $ -0.66   $ &  14 &  $ -0.30   $ &  24 &  $ -0.27   $ &  24 &  $ -0.22   $ &  16 &  $ -0.09   $ &  22 &  $ -0.01  $ &  23   & $ -0.10   $  &  22 &  $ -0.03   $ &  28 &  $ -0.07   $ &  26 & $ -0.02   $ &  17 \\
{Si \sc  ii} &              &     &  $ -0.59  $ &   1 &  $ -0.66   $ &   1 &              &     &  $ -0.18   $ &   1 &  $ -0.21   $ &   1 &              &     &             &       &              &     &              &     &              &     &             &     \\  
{Ca \sc   i} &  $ -1.16   $ &   4 &  $ -0.56  $ &  13 &  $ -0.66   $ &  19 &  $ -0.30   $ &   8 &  $ -0.24   $ &  13 &  $ -0.19   $ &   9 &  $ -0.12   $ &   5 &  $ -0.09  $ &   6   & $ -0.05   $  &   5 &  $ -0.01   $ &  18 &  $ -0.00   $ &  11 & $ +0.08   $ &   7 \\  
{Sc \sc   i} &              &     &  $        $ &     &              &     &              &     &              &     &              &     &  $ -0.24   $ &   1 &  $ -0.31  $ &   1   &              &     &              &     &              &     &             &     \\   
{Sc \sc  ii} &  $ -1.60   $ &   2 &  $ -0.68  $ &   6 &  $ -0.73   $ &   6 &  $ -0.33   $ &   6 &  $ -0.31   $ &   4 &  $ -0.27   $ &   5 &  $ -0.15   $ &   7 &  $ -0.07  $ &   6   & $ -0.05   $  &   5 &  $ -0.04   $ &   8 &  $ -0.06   $ &   6 & $ -0.05   $ &   5 \\  
{Ti \sc   i} &  $ -1.25   $ &   3 &  $ -0.52  $ &  24 &  $ -0.67   $ &  26 &  $ -0.25   $ &  28 &  $ -0.30   $ &  22 &  $ -0.27   $ &  14 &  $ -0.15   $ &  17 &  $ -0.13  $ &  15   & $ -0.08   $  &   8 &  $ -0.04   $ &  27 &  $ -0.09   $ &  20 & $ -0.05   $ &   4 \\   
{Ti \sc  ii} &  $ -1.51   $ &   1 &  $ -0.58  $ &   9 &  $ -0.68   $ &  12 &  $ -0.27   $ &  11 &  $ -0.28   $ &   8 &  $ -0.27   $ &   9 &  $ -0.19   $ &   8 &  $ -0.10  $ &   4   & $ +0.03   $  &   1 &  $ +0.02   $ &  17 &  $ -0.04   $ &   5 & $ +0.06   $ &   3 \\ 
{V  \sc   i} &  $ -1.50   $ &   7 &  $ -0.54  $ &   8 &  $ -0.61   $ &   3 &  $ -0.30   $ &  10 &  $ -0.32   $ &   7 &  $ -0.32   $ &   4 &  $ -0.14   $ &   8 &  $ -0.06  $ &  11   & $ -0.10   $  &   7 &  $ -0.01   $ &   6 &  $ -0.03   $ &   8 & $ -0.09   $ &   6 \\  
{Cr \sc   i} &  $ -1.49   $ &   2 &  $ -0.87  $ &  11 &  $ -0.81   $ &  13 &  $ -0.48   $ &  18 &  $ -0.34   $ &  19 &  $ -0.31   $ &  14 &  $ -0.21   $ &  12 &  $ -0.21  $ &  10   & $ -0.15   $  &   2 &  $ -0.05   $ &  22 &  $ -0.09   $ &  13 & $ -0.02   $ &  12 \\  
{Cr \sc  ii} &              &     &  $ -0.81  $ &   4 &  $ -0.80   $ &   4 &  $ -0.52   $ &   3 &  $ -0.27   $ &   3 &  $ -0.25   $ &   4 &  $ -0.29   $ &   3 &  $ -0.19  $ &   2   & $         $  &     &  $ +0.02   $ &   5 &  $ -0.08   $ &   2 & $ -0.06   $ &   2 \\ 
{Fe \sc   i} &  $ -1.55   $ & 187 &  $ -0.85  $ & 183 &  $ -0.74   $ & 219 &  $ -0.48   $ & 202 &  $ -0.31   $ & 213 &  $ -0.27   $ & 178 &  $ -0.15   $ & 179 &  $ -0.12  $ & 168   & $ -0.11   $  &  92 &  $ -0.05   $ & 260 &  $ -0.06   $ & 216 & $ +0.01   $ & 169 \\  
{Fe \sc  ii} &  $ -1.63   $ &  17 &  $ -0.86  $ &  22 &  $ -0.74   $ &  22 &  $ -0.48   $ &  16 &  $ -0.28   $ &  24 &  $ -0.26   $ &  16 &  $ -0.21   $ &  17 &  $ -0.11  $ &  13   & $ -0.10   $  &   9 &  $ +0.01   $ &  20 &  $ -0.04   $ &  16 & $ +0.01   $ &  17 \\   
{Ni \sc   i} &  $ -1.63   $ &  17 &  $ -0.83  $ &  43 &  $ -0.76   $ &  48 &  $ -0.46   $ &  54 &  $ -0.34   $ &  53 &  $ -0.30   $ &  37 &  $ -0.16   $ &  53 &  $ -0.12  $ &  46   & $ -0.12   $  &  32 &  $ -0.06   $ &  71 &  $ -0.10   $ &  53 & $ -0.03   $ &  35 \\  

\hline
 & \multicolumn{2}{c|}{\opha} & \multicolumn{2}{c|}{\iotahor} & \multicolumn{2}{c|}{\alphamen} & \multicolumn{2}{c|}{\deltaeri}  & \multicolumn{2}{c|}{\betavir} & \multicolumn{2}{c|}{\etaboo} & \multicolumn{2}{c|}{\ksihya} & \multicolumn{2}{c|}{\acena} 
& \multicolumn{2}{c|}{\muara} & \multicolumn{2}{c|}{\acenb} & \multicolumn{2}{c}{\deltapav} &   \\
\hline

{Li \sc   i} &             &     &  $ +1.62   $ &   1 &             &     &  $ +0.37   $  &   1 &  $ +1.06   $ &     1 &              &     &  $ +0.39   $ &   1 &  $ +0.57   $ &   1 &  $ +0.19   $ &   1 &             &     &             &      &             &     \\  
{C  \sc   i} &  $ +0.09  $ &   2 &  $ +0.04   $ &   3 &  $ +0.16  $ &   4 &  $ +0.21   $  &   1 &  $ +0.01   $ &     4 &  $         $ &     &  $ -0.24   $ &   1 &  $ +0.19   $ &   3 &  $ +0.30   $ &   4 &  $ +0.32  $ &   1 &  $ +0.34  $ &   3  &             &     \\ 
{O  \sc   i} &             &     &  $ +0.06   $ &   2 &  $ +0.11  $ &   2 &  $ +0.28   $  &   1 &  $ -0.47   $ &     3 &  $ -0.08   $ &   3 &  $ +0.19   $ &   2 &  $ +0.05   $ &   4 &  $ +0.18   $ &   2 &  $ +0.28  $ &   1 &  $ +0.42  $ &   3  &             &     \\  
{Na \sc   i} &  $ +0.11  $ &   2 &  $ +0.08   $ &   1 &  $ +0.18  $ &   2 &  $ +0.23   $  &   1 &  $ +0.10   $ &     1 &  $ +0.50   $ &   1 &  $ +0.44   $ &   1 &  $ +0.31   $ &   2 &  $ +0.43   $ &   2 &  $ +0.38  $ &   1 &  $ +0.48  $ &   1  &             &     \\  
{Mg \sc   i} &             &     &  $ +0.06   $ &   2 &             &     &               &     &              &       &              &     &              &     &              &     &              &     &             &     &             &      &             &     \\  
{Si \sc   i} &  $ +0.16  $ &  17 &  $ +0.09   $ &  26 &  $ +0.18  $ &  19 &  $ +0.22   $  &  12 &  $ +0.08   $ &     5 &  $ +0.31   $ &  10 &  $ +0.23   $ &  11 &  $ +0.24   $ &  22 &  $ +0.32   $ &  23 &  $ +0.36  $ &  12 &  $ +0.41  $ &  14  &             &     \\  
{Si \sc   i} &             &     &  $ +0.15   $ &   1 &             &     &               &     &              &       &              &     &              &     &  $ +0.16   $ &   1 &  $ +0.32   $ &   1 &  $ +0.43  $ &   1 &             &      &             &     \\  
{Ca \sc   i} &  $ +0.15  $ &   2 &  $ +0.17   $ &  10 &  $ +0.17  $ &   5 &  $ +0.18   $  &   1 &  $ +0.12    $&    10 &  $ +0.21   $ &   2 &  $ +0.22   $ &   1 &  $ +0.20   $ &   5 &  $ +0.28   $ &   5 &  $ +0.31  $ &   2 &  $ +0.37  $ &   4  &             &     \\   
{Sc \sc   i} &  $ +0.04  $ &   1 &  $ +0.06   $ &   1 &  $ +0.10  $ &   1 &               &     &  $ +0.14    $&     1 &              &     &  $ +0.01   $ &   1 &              &     &  $ +0.27   $ &   1 &  $ +0.23  $ &   1 &             &      &             &     \\  
{Sc \sc   i} &  $ +0.06  $ &   8 &  $ +0.09   $ &   4 &  $ +0.19  $ &   7 &  $ +0.14   $  &   2 &  $ +0.15    $&     4 &  $ +0.31   $ &   3 &  $ +0.13   $ &   1 &  $ +0.28   $ &   4 &  $ +0.37   $ &   5 &  $ +0.31  $ &   6 &  $ +0.42  $ &   5  &             &     \\  
{Ti \sc   i} &  $ +0.11  $ &  36 &  $ +0.11   $ &  24 &  $ +0.15  $ &  20 &  $ +0.15   $  &  25 &  $ +0.09    $&    27 &  $ +0.19   $ &   3 &  $ +0.11   $ &  20 &  $ +0.21   $ &  22 &  $ +0.26   $ &  25 &  $ +0.28  $ &  16 &  $ +0.34  $ &  13  &             &     \\  
{Ti \sc   i} &  $ +0.10  $ &  10 &  $ +0.13   $ &  10 &  $ +0.15  $ &  11 &  $ +0.12   $  &   1 &  $ +0.12    $&    13 &  $ +0.22   $ &   1 &  $ +0.20   $ &   4 &  $ +0.21   $ &   7 &  $ +0.30   $ &   9 &  $ +0.37  $ &   1 &  $ +0.28  $ &   3  &             &     \\  
{V  \sc   i} &  $ +0.20  $ &  17 &  $ +0.13   $ &   9 &  $ +0.20  $ &   5 &  $ +0.24   $  &   7 &  $ +0.09    $&     1 &  $ +0.22   $ &   7 &  $ +0.20   $ &   9 &  $ +0.24   $ &  10 &  $ +0.30   $ &   9 &  $ +0.47  $ &  16 &  $ +0.45  $ &  10  &             &     \\  
{Cr \sc   i} &  $ +0.09  $ &  13 &  $ +0.14   $ &  14 &  $ +0.14  $ &  12 &  $ +0.11   $  &   7 &  $ +0.11    $&    17 &  $ +0.24   $ &   4 &  $ +0.16   $ &   7 &  $ +0.21   $ &  11 &  $ +0.28   $ &  11 &  $ +0.26  $ &  12 &  $ +0.36  $ &   8  &             &     \\  
{Cr \sc   i} &  $ +0.08  $ &   4 &  $ +0.15   $ &   1 &  $ +0.09  $ &   4 &               &     &  $ +0.09    $&     5 &  $         $ &     &  $ +0.14   $ &   1 &  $ +0.18   $ &   2 &  $ +0.27   $ &   5 &             &     &             &      &             &     \\   
{Fe \sc   i} &  $ +0.11  $ & 217 &  $ +0.14   $ & 206 &  $ +0.15  $ & 166 &  $ +0.16   $  & 116 &  $ +0.11    $&   188 &  $ +0.23   $ &  70 &  $ +0.20   $ &  99 &  $ +0.22   $ & 184 &  $ +0.29   $ & 186 &  $ +0.31  $ & 129 &  $ +0.38  $ & 126  &             &     \\   
{Fe \sc   i} &  $ +0.12  $ &  13 &  $ +0.15   $ &  20 &  $ +0.15  $ &  21 &  $ +0.15   $  &   7 &  $ +0.12    $&    18 &  $ +0.24   $ &   8 &  $ +0.21   $ &  10 &  $ +0.22   $ &  19 &  $ +0.32   $ &  16 &  $ +0.30  $ &   8 &  $ +0.38  $ &  12  &             &     \\   
{Ni \sc   i} &  $ +0.12  $ &  61 &  $ +0.11   $ &  58 &  $ +0.18  $ &  55 &  $ +0.21   $  &  32 &  $ +0.13    $&    62 &  $ +0.27   $ &  22 &  $ +0.18   $ &  24 &  $ +0.26   $ &  60 &  $ +0.34   $ &  56 &  $ +0.38  $ &  33 &  $ +0.43  $ &  34  &             &     \\  

\hline

\end{tabular}
\end{footnotesize}
\end{table*}

\section{Conclusions and future outlook}

We have determined the fundamental parameters of 23 bright solar-type stars: mass, radius, and luminosity. 
Our goal was to assess the absolute accuracy of indirect techniques 
by comparing our results with direct techniques that are only weakly model-dependent.
The adopted direct techniques used interferometric data, 
bolometric fluxes and parallaxes, or orbits of binary stars,
and could be applied to 10 of the stars in the sample.
The indirect methods used asteroseismic data, 
spectroscopic data, and \str\ photometry.
We also presented a detailed spectroscopic analysis of high-quality spectra. 
This included the determination of \teff, chemical composition, 
surface gravity, and projected rotational velocity. 
We compared the determined parameters with results from the literature 
that use similar spectroscopic methods, and found good agreement except for a few cases.
In summary, indirect mass and radius estimates give good results, 
with some evidence for systematic errors in the luminosity of cool stars,
while spectroscopic \teff\ values need a slight adjustment. 
For future analyses, we conclude that from spectroscopic analysis 
of a high-quality spectrum, 
\teff\ can be determined to 80~K, \logg\ to 0.08~dex, and \feh\ to 0.07~dex.
Similar conclusions, based on larger spectroscopic
data sets, were reached by \cite{fuhrmann04} and \cite{valenti05}.

We have determined a homogeneous set of parameters for 23 solar-type stars
that will be valuable for future asteroseismic campaigns.
Observations have already been carried out on 22 of the 23 stars
and have shown that the stars are indeed oscillating. 
In the near future several of the stars analysed here will 
be targets for the Stellar Observations Network Group 
(\song; \citealt{grundahl08, grundahl09}). 
Such lists of bright targets for asteroseismology have 
previously been given by \cite{bedding96} and \cite{pijpers03}. 
Although these are extremely useful for selecting targets, 
they may be of limited use when doing detailed asteroseismic analyses. 
For example, the \cite{bedding96} study pre-dates the \hipp\ catalogue, while
the parameters tabulated by \cite{pijpers03} 
were determined from multiple studies using different methods and 
varying quality of observations, \eg\ 
18 different papers cited for the effective temperature of 40 stars.


Looking ahead, improvements in \fbol\ determinations may soon be possible.
The accuracy is limited by the accuracy of the
stellar flux scale, which is currently based on the \cite{hayes75} 
absolute flux measurements of Vega. This gives a minimum
uncertainty of around 2\percent\ on \fbol. 
The internal uncertainties in spectrophotometry of the
individual stars contribute at least 1\percent\ to the
final error determination. This limits direct \teff\ determinations to no
better than around $\pm$50~K for solar-type stars.
The ACCESS project 
(Absolute Color Calibration Experiment for Standard Stars; \citealt{kaiser08}) 
should determine the absolute flux scale to better than 1\percent. 
This, together with improved stellar
spectrophotometry \citep{adelman07}, will enable \fbol\ 
(and hence \teff) to be determined to higher accuracy.

As a future outlook on the subject of accurate stellar parameters, 
we note that while interferometric measurements 
are available for only ten of the stars in our sample, 
all of the targets should now be possible to observe with the recent 
improvements in intrumentation like the \pavo\ instrument at 
\chara\ and \susi\ \citep{ireland08} or \amber\ at \vlti\ \citep{amber07}.
We strongly encourage such measurements to improve the
calibration of secondary methods to characterize solar-type stars.
This will be invaluable to improve the modelling efforts using asteroseismology
of more distant targets of the \corot\ and \kepler\ missions.

\section*{Acknowledgments}

This project was supported by 
the Australian and Danish Research Councils.
We made use of the SIMBAD database, operated at CDS, Strasbourg, France.
This research has made use of NASA's Astrophysics Data System Bibliographic Services.
We used spectra from HARPS obtained from the ESO archive
under programme IDs 60.A-9036, 073.D-0590, 073.D-0578, 073.D-0590, 074.D-0380, 075.D-0760, 
076.D-0103, 077.D-0498, 077.C-0530, 077.D-0720, 078.D-0067, 078.D-0492, 078.C-0233, 079.D-0466, 
079.C-0681, 080.C-0712, and 081.D-0531.


\bibliography{bruntt_solar}

\newpage

\appendix

\section{Comparison with other spectroscopic studies} 
\label{app:tables}

%
%
\begin{table*}
 \centering
  \caption{Comparison of effective temperatures from this study (\str, \vwa) with values found in the literature 
as identified by the first author name. 
\label{tab:teff}}
   \begin{tabular}{llllllll}
   \hline

       Star & \str        & \vwa         & Fuhrmann    & Sousa       & Soubiran    & Santos      &  Valenti \\ \hline
   \betahyi &$5870\pm  70$&$5830\pm  80$&$           $&$           $&$           $&$           $&$5873\pm  44$\\
   \taucet  &$5410\pm  70$&$5330\pm  80$&$5373\pm  80$&$5310\pm  17$&$5264\pm 100$&$5344\pm  29$&$5283\pm  44$\\
   \iotahor &$6110\pm  70$&$6120\pm  80$&$           $&$           $&$           $&$6252\pm  53$&$6097\pm  44$\\

    \alphafor&$6105\pm  60$&$6055\pm  80$&$           $&$           $&$           $&$6275\pm  57$&$           $\\
    \deltaeri&             &$5055\pm  80$&$5044\pm  80$&$5150\pm  51$&$           $&$5074\pm  60$&$           $\\
    \alphamen&$5580\pm  70$&$5610\pm  80$&$           $&$           $&$           $&$5594\pm  36$&$5587\pm  44$\\
    \procyon &$6595\pm  70$&$6525\pm  80$&$6470\pm  80$&$           $&$           $&$           $&$           $\\
    \betavir &$6150\pm  70$&$6090\pm  80$&$6085\pm  80$&$           $&$6109\pm  75$&$           $&$           $\\
    \etaboo  &$6025\pm  60$&$6070\pm  80$&$6023\pm  70$&$           $&$           $&$           $&$           $\\
   \gammaser &$6245\pm  70$&$6155\pm  80$&$6254\pm  80$&$           $&$           $&$           $&$           $\\
    \muara   &$5690\pm  70$&$5705\pm  80$&$           $&$5780\pm  25$&$           $&$5798\pm  33$&$           $\\
    \opha    &$           $&$5340\pm  80$&$5310\pm  80$&$           $&$5158\pm 250$&$           $&$           $\\
    \etaser  &$           $&$4890\pm  80$&$4921\pm  80$&$           $&$4859\pm 250$&$           $&$           $\\
    \betaaql &             &$5070\pm  80$&$5110\pm  90$&$           $&$5041\pm 200$&$           $&$           $\\
    \deltapav&$5540\pm  70$&$5590\pm  80$&$           $&$5604\pm  38$&$           $&$           $&$           $\\

\hline
\end{tabular}
\end{table*}
%

%
%
\begin{table*}
 \centering
  \caption{Comparison of the surface gravity from \vwa\ with values found in the literature 
as identified by the first author name. 
\label{tab:loggcomp}}
   \begin{tabular}{lllllll}
   \hline

       Star &  \vwa          & Fuhrmann      & Sousa         & Soubiran      & Santos        & Valenti   \\ \hline

   \betahyi &$ 3.84\pm 0.08$&$             $&$             $&$             $&$             $&$ 4.08\pm 0.06$\\
   \taucet  &$ 4.46\pm 0.08$&$ 4.54\pm 0.10$&$ 4.44\pm 0.03$&$ 4.36\pm 0.30$&$ 4.57\pm 0.09$&$ 4.59\pm 0.06$\\
   \iotahor &$ 4.40\pm 0.08$&$             $&$             $&$             $&$ 4.61\pm 0.16$&$ 4.34\pm 0.06$\\
    \alphafor&$ 3.80\pm 0.08$&$             $&$             $&$             $&$ 4.40\pm 0.37$&$             $\\
    \deltaeri&$ 3.77\pm 0.08$&$ 3.84\pm 0.10$&$ 3.89\pm 0.08$&$             $&$ 3.77\pm 0.16$&$             $\\
    \alphamen&$ 4.43\pm 0.08$&$             $&$             $&$             $&$ 4.41\pm 0.09$&$ 4.50\pm 0.06$\\
    \procyon &$ 3.89\pm 0.08$&$ 4.01\pm 0.10$&$             $&$             $&$             $&$             $\\
    \betavir &$ 3.98\pm 0.08$&$ 4.04\pm 0.10$&$             $&$ 4.20\pm 0.30$&$             $&$             $\\
    \etaboo  &$ 3.90\pm 0.08$&$ 3.76\pm 0.10$&$             $&$             $&$             $&$             $\\
   \gammaser &$ 4.01\pm 0.08$&$ 4.02\pm 0.10$&$             $&$             $&$             $&$             $\\
    \muara   &$ 4.14\pm 0.08$&$             $&$ 4.27\pm 0.04$&$             $&$ 4.31\pm 0.08$&$             $\\
    \opha    &$ 4.57\pm 0.08$&$ 4.57\pm 0.10$&$             $&$ 4.67\pm 0.30$&$             $&$             $\\
    \etaser  &$ 2.96\pm 0.08$&$ 2.99\pm 0.10$&$             $&$ 3.13\pm 0.30$&$             $&$             $\\
    \betaaql &$ 3.55\pm 0.08$&$ 3.60\pm 0.10$&$             $&$ 3.04\pm 0.30$&$             $&$             $\\
    \deltapav&$ 4.32\pm 0.08$&$             $&$ 4.26\pm 0.06$&$             $&$             $&$             $\\

\hline
\end{tabular}
\end{table*}
%

%
%
\begin{table*}
 \centering
  \caption{Comparison of the metallicity from \vwa\ with values found in the literature 
as identified by the first author name. 
\label{tab:feh}}
   \begin{tabular}{llllllll}
   \hline

       Star & \str          & \vwa           & Fuhrmann      & Sousa         & Soubiran      & Santos        & Valenti\\ \hline
   \betahyi &$-0.04\pm 0.10$&$-0.10\pm 0.07$&$             $&$             $&$             $&$             $&$-0.09\pm 0.03$\\
   \taucet  &$-0.42\pm 0.10$&$-0.48\pm 0.07$&$-0.53\pm 0.07$&$-0.52\pm 0.01$&$-0.50\pm 0.15$&$-0.52\pm 0.04$&$-0.36\pm 0.03$\\
   \iotahor &$-0.00\pm 0.10$&$+0.15\pm 0.07$&$             $&$             $&$             $&$+0.26\pm 0.06$&$+0.09\pm 0.03$\\
    \alphafor&$-0.16\pm 0.10$&$-0.28\pm 0.07$&$             $&$             $&$             $&$-0.19\pm 0.06$&$             $\\
    \deltaeri&               &$+0.15\pm 0.07$&$+0.12\pm 0.07$&$+0.13\pm 0.04$&$             $&$+0.13\pm 0.08$&$             $\\
    \alphamen&$+0.07\pm 0.10$&$+0.15\pm 0.07$&$             $&$             $&$             $&$+0.10\pm 0.05$&$+0.05\pm 0.03$\\
    \procyon &$+0.02\pm 0.10$&$+0.01\pm 0.07$&$-0.01\pm 0.07$&$             $&$             $&$             $&$             $\\
    \betavir &$+0.10\pm 0.10$&$+0.12\pm 0.07$&$+0.14\pm 0.09$&$             $&$+0.17\pm 0.15$&$             $&$             $\\
    \etaboo  &$+0.27\pm 0.10$&$+0.24\pm 0.07$&$+0.28\pm 0.07$&$             $&$             $&$             $&$             $\\
   \gammaser &$-0.19\pm 0.10$&$-0.26\pm 0.07$&$-0.19\pm 0.07$&$             $&$             $&$             $&$             $\\
    \muara   &$+0.19\pm 0.10$&$+0.32\pm 0.07$&$             $&$+0.30\pm 0.02$&$             $&$+0.32\pm 0.04$&$             $\\
    \opha    &$             $&$+0.12\pm 0.07$&$+0.06\pm 0.07$&$             $&$+0.03\pm 0.15$&$             $&$             $\\
    \etaser  &$             $&$-0.11\pm 0.07$&$-0.21\pm 0.07$&$             $&$-0.19\pm 0.15$&$             $&$             $\\
    \betaaql &$             $&$-0.21\pm 0.07$&$-0.17\pm 0.07$&$             $&$-0.04\pm 0.15$&$             $&$             $\\
    \deltapav&$+0.33\pm 0.10$&$+0.38\pm 0.07$&$             $&$+0.33\pm 0.03$&$             $&$             $&$             $\\

\hline
\end{tabular}
\end{table*}
%

In Tables~\ref{tab:teff}--\ref{tab:feh} we compare the atmospheric parameters 
from spectroscopic analysis (\vwa) and \str\ calibrations with values from the literature. 
Since we compare directly with similar spectroscopic techniques,
we have not applied the $-40$~K offset to the \teff\ from \vwa\ (see Sect.~\ref{sec:teff}).
Each column is identified by the first author, \ie\ \cite{fuhrmann98, fuhrmann04, fuhrmann08},
\cite{sousa08}, \cite{soubiran98}, \cite{santos04, santos05}, and \cite{valenti05}. 
We also list \teff\ and \feh\ using the \str\ calibration by \cite{holmberg07}.

\section[]{Projected rotational velocities}
\label{ap:vsini}

In Table~\ref{tab:vsini} we list \vsini\ and \vmacro\ determined by the
technique described in Sect.~\ref{sec:vsini}.
Our results are in general agreement with similar investigations 
in the literature, identified by the first author in the Table: 
\cite{saar97}, \cite{valenti05}, \cite{reiners03}, \cite{dravins93}, and \cite{allende02}. 
Only \cite{saar97} determined both \vsini\ and \vmacro\ with an approach similar to ours.
\cite{dravins93} and \cite{allende02} used very high resolution spectra ($R\simeq200\,000$)
and compared their observed line profiles with 3D hydrodynamical simulations.
It is encouraging that our method also matches the two later studies for
\betahyi\ and \procyon\ very well.

%
\begin{table*}
 \centering
  \caption{The determined projected rotational velocity and macroturbulence from
this study is compared with five other studies from the literature identified by
the first author name. The unit is \kms\ for all measured parameters.
\label{tab:vsini}}
   \begin{tabular}{lrr rrrrr rr}
   \hline

          &            
\multicolumn{1}{c}{This study} &
\multicolumn{1}{c}{This study} &
\multicolumn{1}{c}{Saar}       & 
\multicolumn{1}{c}{Saar}       & 
                                   \multicolumn{1}{c}{Valenti} &  \multicolumn{1}{c}{Reiners}       & \multicolumn{1}{c}{Dravins}      & \multicolumn{1}{c}{Allende} \\
Star      & \multicolumn{1}{c}{\vsini}       &       \multicolumn{1}{c}{\vmacroo}   &     \multicolumn{1}{c}{\vsini}  & \multicolumn{1}{c}{\vmacroo}        & \multicolumn{1}{c}{\vsini} &  \multicolumn{1}{c}{\vsini}        &  \multicolumn{1}{c}{\vsini}     & \multicolumn{1}{c}{\vsini}  \\ \hline
   Sun    &   $   1.3\pm  0.6$ & $ 2.4\pm  0.6$& $2.3\pm0.6$ & $1.9\pm0.8$     &        & $            $ &              &         \\
 \betahyi &   $   2.7\pm  0.6$ & $ 2.9\pm  0.6$&             &                 & $4.0 $ & $  3.3\pm0.3 $ & $2\pm1 $  \\ 
 \taucet  &   $   0.7\pm  0.6$ & $ 0.9\pm  0.6$& $2.1\pm0.4$ & $0.4\pm0.4$     & $1.3 $ & $ <1.8\pm0.1 $ &           \\ 
 \iotahor &   $   5.3\pm  0.6$ & $ 3.3\pm  0.6$&             &                 & $6.5 $ & $  4.2\pm0.6 $ &           \\ 
  \alphafor&   $   3.9\pm  0.6$ & $ 3.7\pm  0.6$&             &                 &        & $  4.0\pm0.7 $ &              &         \\
  \deltaeri&   $   0.7\pm  0.6$ & $ 0.9\pm  0.6$&             &                 &        & $ <2.3\pm0.5 $ &              &         \\
  \alphamen&   $   0.6\pm  0.6$ & $ 1.0\pm  0.6$&             &                 &        &                &              &         \\
  \procyon &   $   2.8\pm  0.6$ & $ 4.6\pm  0.6$&             &                 & $5.7 $ &                &              & $2.7$   \\ 
  \pup     &   $   1.6\pm  0.6$ & $ 1.9\pm  0.6$&             &                 &        &                &              &         \\
  \ksihya  &   $   2.4\pm  0.6$ & $ 3.8\pm  0.6$&             &                 &        &                &              &         \\
  \betavir &   $   2.0\pm  0.6$ & $ 3.6\pm  0.6$&             &                 &        &                &              &         \\
  \etaboo  &   $  11.9\pm  0.6$ & $ 5.3\pm  0.6$&             &                 &        & $ 13.5\pm1.3 $ &              &         \\
  \acena   &   $   1.9\pm  0.6$ & $ 2.3\pm  0.6$& $2.6\pm0.9$ & $2.7\pm0.7$     & $2.3 $ & $            $ &              &         \\
  \acenb   &   $   1.0\pm  0.6$ & $ 0.8\pm  0.6$& $1.0\pm0.8$ & $1.1\pm0.8$     & $0.9 $ & $            $ &              &         \\ 
 \hdcarrier&   $   5.2\pm  0.6$ & $ 4.1\pm  0.6$&             &                 &        & $  5.0\pm0.2 $ &              &         \\
 \gammaser &   $  10.2\pm  0.6$ & $ 3.9\pm  0.6$&             &                 &        & $ 10.0\pm1.3 $ &              &         \\
  \muara   &   $   1.4\pm  0.6$ & $ 2.6\pm  0.6$&             &                 &        & $  3.8\pm0.2 $ &              &         \\
  \opha    &   $   0.9\pm  0.6$ & $ 1.5\pm  0.6$&             &                 &        &                &              &         \\
  \etaser  &   $   1.1\pm  0.6$ & $ 2.1\pm  0.6$&             &                 &        &                &              &         \\
  \betaaql &   $   0.9\pm  0.6$ & $ 2.1\pm  0.6$&             &                 &        & $ <2.3\pm0.3 $ &              &         \\
  \deltapav&   $   1.0\pm  0.6$ & $ 1.7\pm  0.6$&             &                 &        & $  3.2\pm0.2 $ &              &         \\
  \gammapav&   $   1.8\pm  0.6$ & $ 2.0\pm  0.6$&             &                 &        & $  2.4\pm0.5 $ &              &         \\
  \taupsa  &   $  13.6\pm  0.6$ & $ 4.1\pm  0.6$&             &                 &        & $ 13.6\pm0.2 $ &              &         \\
  \nuind   &   $   1.1\pm  0.6$ & $ 1.4\pm  0.6$&             &                 &        &                &              &         \\
\hline

 \end{tabular}
 \end{table*}
%

\bsp 

\label{lastpage}

\end{document}